\documentstyle[12pt,twoside]{article}
\begin{document}

\newif\iffigs\figsfalse
\figstrue

\iffigs
  \input epsf
\else
  \message{No figures will be included. See TeX file for more
information.}
\fi

\thispagestyle{empty}
\begin{flushright}
MPI-PhT/96-116
\end{flushright}
\bigskip\bigskip\bigskip\begin{center}
{\LARGE {A determination of the mass gap}}
\vskip 7pt
{\LARGE {in the O($n$) $\sigma$ model}}
\end{center}  
\vskip 1.0truecm
\centerline{Dong-Shin Shin}
\vskip5mm
\centerline{Max-Planck-Institut f\"ur
 Physik}
\centerline{ -- Werner-Heisenberg-Institut -- }
\centerline{F\"ohringer Ring 6, 80805 Munich, Germany}
\vskip 2cm
\bigskip \nopagebreak \begin{abstract}
\noindent
We calculate the finite volume mass gap $M(L)$ at 3-loop level in 
the non-linear O($n$) $\sigma$-model in two dimensions in small volumes.
By applying the Monte Carlo measurements of the 
running coupling $\bar g^2(L)=2nM(L)L/(n-1)$ 
by L\"uscher, Weisz and Wolff \cite{LuWeWo}
measured in units of the physical mass gap $m$, 
the result is used to determine $m$ in units of the
$\Lambda$-parameter in the O(3) and O(4) models.
Our determinations show good agreement with those 
by Hasenfratz, Maggiore and Niedermayer \cite{HaMaNi,HaNi} in both models.
We note that this manuscript has been revised in our paper hep-lat/9810025 
by using the corrected four-loop $\beta$-function on the lattice.

\end{abstract}
\vskip 1.5cm

\newpage\setcounter{page}1

\def\eqb{\begin{equation}}
\def\eqe{\end{equation}}
\def\dib{\begin{displaymath}}
\def\die{\end{displaymath}}
\def\eqnb{\begin{eqnarray}}
\def\eqne{\end{eqnarray}}
\def\eqsb{\begin{eqnarray*}}
\def\eqse{\end{eqnarray*}}

\section{Introduction}

The non-linear $\mbox{O}(n)$ $\sigma$-model which describes an
$n$-component spin field has thus far found many applications 
in theoretical physics. In condensed matter physics 
it has been investigated 
to study, e.g. ferromagnets. In elementary particle physics,
on the other hand,
the model in two dimensions shares with QCD many common properties;
on the quantum level it is like QCD
renormalizable \cite{BrZJLeG},
asymptotically free according to perturbation theory 
\cite{Pol,BrZJaf,BaLeSh}
and thought to have a mass gap $m$.
In addition, the O(3) model has instanton solutions.
Further, the $\sigma$-model has an additional advantage:
because of low dimensionality it is simple to simulate numerically.
Therefore, in many cases the model can be used
for tests of new ideas in lattice theories.

One of the most interesting properties of the $\sigma$-model is
the existence of an infinite set of conserved non-local
quantum charges \cite{Lu1,Lu2,Bu1,Bu2}. 
From these, one can exclude particle production
and also derive the ``factorization equations'' directly.
Using these properties, one can then determine 
the $S$-matrix exactly up to CDD-ambiguity \cite{CDD},
provided the particle spectrum of the model is known \cite{Za1,Za2}.
Furthermore, the CDD-ambiguity can also be restricted
with help of the assumption that there are no bound states.
The absence of the bound states was shown
in the $1/n$-expansion to the order of $1/n$ \cite{Za1}.
Further, the exact $S$-matrix constructed in this way
was confirmed by the same method to the second order in $1/n$ \cite{BKKW}.

Since there is only one scale in the model, 
one can, in principle, express all physical quantities 
with help of this scale.
In particular, the mass gap of the theory is proportional to the
$\Lambda$-parameter.
This situation is similar to what happens in QCD with massless quarks
where the $\Lambda$-parameter sets the scale for scale violations,
for example, in deep inelastic scattering.

Using the exact $S$-matrix, a few years ago 
Hasenfratz, Maggiore and Niedermayer succeeded in an analytic
determination of the ratio $m/\Lambda$ in the $\mbox{O}(n)$ $\sigma$-model
\cite{HaMaNi,HaNi}.
In this derivation, however, various assumptions 
including the thermodynamical Bethe ansatz are made,
which are plausible, 
for which, however, the rigorous proofs are still lacking.
An independent calculation of $m/\Lambda$ is therefore desirable.

In the literature \cite{LuWeWo},
L\"uscher, Weisz and Wolff estimated the ratio $m/\Lambda$
by measuring the coupling $\bar g^2(L)=2nM(L)L/(n-1)$ 
running with the volume $L$ of the system
in units of the physical mass gap $m$[=$M(\infty)$]
by means of Monte Carlo simulations
and applying the measurements to the perturbative
expression of $\Lambda$-parameter in this scheme.
Although their determination of $m/\Lambda$
where $\Lambda$ was evaluated up to three loops in perturbation theory
reproduces that of Hasenfratz et. al quite well,
the agreement of two results was still not completely undisputable.
In order to determine $m/\Lambda$ more precisely,
we decided to evaluate the $\Lambda$-parameter to one loop order higher.
This requires the calculation of 
the finite volume mass gap $M(L)$ at 3-loop level.

Our work is also interesting in another respect.
It is generally accepted that
the $\mbox{O}(n)$ $\sigma$-model is asymptotically free.
This assumption was, however, questioned by Patrascioiu and Seiler
\cite{PS,PSI,PSII}.
Since in the determination 
of $m/\Lambda$ by Hasenfratz et al. the validity
of the asymptotic freedom was required,
the confirmation of their result
through our independent determination will, at the same time,
support the correctness of asymptotic freedom.

The work is arranged as follows.
In the next chapter, we present our 3-loop
calculations of the finite volume mass gap $M(L)$.
By computing the spin-spin correlation function, 
we first evaluate the mass gap on a lattice.
We then take the continuum limit and finally convert it to the MS-scheme.
In chapter~\ref{cha3}, we apply the result to 
the determination of mass gap $m$ in units of the $\Lambda$-parameter
in the O(3) and O(4) models.
Two approaches will be considered. 
First, we introduce the mass shift
$\delta_0=[M(L)-m]/m$ which is known for $L\to\infty$ and match this with 
the perturbative $M(L)$ \cite{Lup,lueschercargese,FlPe}.
As a second approach, 
we determine $m$ by applying the Monte Carlo measurements
of the running coupling $\bar g^2(L)$
to the perturbative $\Lambda$-parameter \cite{LuWeWo}.
Finally, the Feynman diagrams and some technical details are presented
in the appendices.

\section{Calculation of mass gap at 3-loop level}\label{cha2}

\subsection{Mass gap on a lattice}

We consider an $n$-component spin field
$q^i(x)\,\,(i=1,\cdots,n)$ with unit length $q(x)^2 = 1$ on a
two-dimensional finite lattice\footnote{We set the 
lattice constant $a$ to 1.}
\begin{equation} 
\Lambda = \bigg\{x\in {\bf Z}^2 | -T\leq x_0 \leq T ;
\,\,x_1 \equiv x_1 + L\bigg\},
\end{equation}
where $T$ and $L$ are fixed integers with $T,L\geq 2$. 
We work with periodic boundary conditions
in space direction and,
for reasons which will become clear below,
free boundary conditions in time direction.

The action is given by
\begin{equation}\label{oldaction}
S = \frac{n}{2f_0}\sum_{x_1=1}^{L}
\bigg\{\sum_{x_0=-T}^{T-1}\partial_0 q(x)\cdot\partial_0 q(x) + 
\sum_{x_0=-T}^{T}\partial_1 q(x)\cdot\partial_1 q(x)\bigg\},
\end{equation}
where $f_0$ denotes the bare 
coupling constant.
The forward and backward lattice derivatives are defined by
\begin{eqnarray*}
\partial_{\mu}q(x) & = & q(x+\hat \mu) - q(x) \\
\partial_{\mu}^{\ast}q(x) & = & q(x) - q(x-\hat\mu).
\end{eqnarray*}
$\hat\mu$ is the unit vector to the positive $\mu$-direction ($\mu=0,1$).
The expectation values of the 
observables can be calculated by the formula
\begin{equation}\label{expectvalue}
\langle {\cal O}\rangle = \frac{1}{Z}\int\prod_{x\in \Lambda}
\Big[d^nq(x)\delta(q(x)^2-1)\Big]{\cal O} e^{-S},
\end{equation}
where the partition function $Z$ is such that $\langle 1\rangle = 1$.

Concerning the energy spectrum, there is a following possibility
for the calculation of the mass gap $M(L)$ in finite volume for $f_0\to 0$.
We calculate the expectation value of the spatially averaged 
spin-spin correlation function in the limit $T\to\infty$ ($\tau >0$):
\begin{equation}\label{correl}
C(\tau) = \lim_{T\to\infty}\frac{1}{L^2}\sum_{x_1=1}^{L}\sum_{y_1=1}^{L}
\langle q(x)\cdot q(y)\rangle |_{x_0=-y_0=\tau} .
\end{equation}
If the bare coupling $f_0$ is fixed, the correlation function
$C(\tau)$ converges to 
the vacuum expectation value of the 2-point function for 
spin field operators at large $T$  
independently of whether one chooses free or periodic 
boundary conditions in the time direction.
In perturbation theory, we are, however, first expanding 
in powers of the coupling $f_0$ and then let $T$ go to infinity.
Therefore, we can not, in general, have such a converging behavior,
and one requires a proper choice 
of boundary conditions to get the desired vacuum expectation value.

In free boundary conditions, we are projecting on states which are
O($n$) invariant at large times.
The energies $\varepsilon (f_0)$ 
in these states, on the other hand, have the property
\begin{equation} \label{posen}
\lim_{f_0 \to 0} \Big\{\varepsilon (f_0) - \varepsilon_0(f_0)\Big\} > 0 ,
\end{equation}
where $\varepsilon_0(f_0)$ denotes the ground state energy.
Except for the ground state, all these states are therefore exponentially 
suppressed at large $T$ and we arrive at the desired vacuum expectation value
even though we are first expanding in powers of $f_0$ \cite{LuWeWo}.

After having taken $T$ to infinity so that $C(\tau)$ been reduced to 
the vacuum expectation value of the product of two spin field operators,
only the vector intermediate states contribute.
The energy $\varepsilon_1(f_0)$ of the first excited state of this system
has the property
\begin{equation}\label{lowestexen}
\varepsilon_1(f_0)-\varepsilon_0(f_0) = {\cal O}(f_0)
\hspace{0.5cm} \mbox{for} \hspace{0.3cm} f_0\to 0 .
\end{equation}
For the energies of the other higher excited states, 
eq.(\ref{posen}) is valid and 
their contribution to the spin correlation function hence vanishes
exponentially at large $\tau$.

From this consideration,
the mass gap $M(L)$ which is defined by the l.h.s of eq.(\ref{lowestexen})
can be calculated by
\begin{equation}\label{endif}
M(L) = - \lim_{\tau\to\infty}\frac12 
\frac{\partial}{\partial\tau} \ln C(\tau).
\end{equation}
If one expands $C(\tau)$ in perturbation theory for $f_0\to 0$,
the mass gap has the general form
\begin{equation}\label{massgapexp}
M(L) = \frac{1}{2L}\sum_{\nu=1}^{\infty}f_0^{\nu}\Delta^{(\nu)}.
\end{equation}
$\Delta^{(\nu)}$ can thus be determined 
by calculating the 2-point function [eq.(\ref{correl})]
with the mentioned boundary conditions in perturbation theory.

The calculations up to the third order in $f_0$
were already done by L\"uscher and Weisz \cite{LWU}.
Their results read
\begin{eqnarray}\label{condelta1}
\Delta^{(1)} &=&  \Delta_0^{(1)} \\
\Delta^{(2)} &=&  \Delta_0^{(2)} + \Delta_1^{(2)}[\ln L] \\\label{condelta3}
\Delta^{(3)} &=&  \Delta_0^{(3)}+\Delta_1^{(3)}[\ln L]
                 +\Delta_2^{(3)}[\ln L]^2
\end{eqnarray}
where the coefficients are given by 
\begin{eqnarray}
\Delta_0^{(1)} &=& \frac{n-1}{n} \\
\Delta_0^{(2)} &=& -\frac{(n-1)^2}{2\pi n^2}
  \bigg\{\ln\frac{\pi}{\sqrt{2}}-\gamma\bigg\}
  +\frac{n-1}{2\pi n^2}
  \bigg\{\ln\frac{\pi}{\sqrt{2}}+\frac{\pi}{2}-\gamma\bigg\} \\
\Delta_1^{(2)} &=& \frac{(n-1)^2}{2\pi n^2}-\frac{n-1}{2\pi n^2} \\
\Delta_0^{(3)} &=& \frac{n-1}{n^3}\Big\{k_0+(n-1)k_1+(n-1)^2k_2\Big\} \\
\Delta_1^{(3)} &=& -\frac{(n-1)^3}{(2\pi)^2n^3}\bigg\{
  2\ln\frac{\pi}{\sqrt{2}}-2\gamma\bigg\}\nonumber\\ & &
  +\frac{(n-1)^2}{(2\pi)^2n^3}\bigg\{
  4\ln\frac{\pi}{\sqrt{2}}+\pi-4\gamma +1\bigg\} \nonumber\\ & &
  -\frac{n-1}{(2\pi)^2n^3}\bigg\{
  2\ln\frac{\pi}{\sqrt{2}}+\pi-2\gamma +1\bigg\} \\
\Delta_2^{(3)} &=& \frac{(n-1)^3}{(2\pi)^2n^3}-2\frac{(n-1)^2}{(2\pi)^2n^3}
                  +\frac{n-1}{(2\pi)^2n^3} 
\end{eqnarray}
with
\begin{eqnarray}
k_0 &=& 0.111\,419\,436\,811\,28(1) \\
k_1 &=& 0.001\,927\,404\,148\,69(1) \\
k_2 &=& \frac{1}{(2\pi)^2}\bigg(\ln\frac{\pi}{\sqrt{2}}-\gamma\bigg)^2
\end{eqnarray}
$\gamma$ denotes the Euler constant ($\gamma=0.577216\cdots$).

In order to determine the mass gap in the fourth order,
we first evaluate $\Delta^{(4)}$ 
on finite lattices and then extrapolate to continuum
limit, i.e., $L/a\to\infty$.
The 2- and 3-loop diagrams contributing to $\Delta^{(4)}$
are illustrated in figures~(\ref{fd2}) and (\ref{fd3})
which are presented in appendix~A due to the very large number of the diagrams.
In carring out Wick's contractions from them,
we used the symbolic language MATHEMATICA.

We calculated the generated terms numerically.
In the numerical computations, we take $T$ and $\tau$ large, but finite.
The corrections to the limit $T,\tau\to\infty$ are here exponentially
suppressed by the order of 
$e^{-\frac{4\pi}{L}2\tau}$ and $e^{-\frac{4\pi}{L}(T-\tau)}$.
We therefore achieve the best approxiamation to the limit $T,\tau\to\infty$
by keeping $T\simeq 3\tau$.

The numerical work, however, turned out to be very complicated
due to the problems caused by rounding errors and running time
which increases with $L$ very quickly (in the order of $L^6$). 
In addition, problems appear from the fact that 
on the 3-loop level on which we are working
there are extremely many terms to treat.
One needs therefore  a very efficient computer program 
which requires, among others,
the free propagator running fast without significant loss of precision.
In this way, we succeeded in evaluating all diagrams up to
$L=20$ with good enough precision and reasonable running time
(for $L=20$ we needed the CPU time of around 7 days on the 
IBM RISC 6000/32H), 
so that we could extrapolate to the continuum limit
to get our desired results.

With regard to the extrapolation to the continuum limit,
we note that in this limit $\Delta^{(4)}$ has the general form
\begin{equation}\label{condelta4}
\Delta^{(4)} = \Delta_0^{(4)}+\Delta_1^{(4)}(\ln L)
                 +\Delta_2^{(4)}(\ln L)^2+\Delta_3^{(4)}(\ln L)^3
\end{equation}
up to terms of the order $L^{-2}(\ln L)^{3}$ where the coefficients
$\Delta_0^{(4)},\cdots,\Delta_3^{(4)}$ are independent of $L$.
All of these coefficients except $\Delta_0^{(4)}$
can be derived with help of the  
renormalization group equation and the results from the 
calculations of the mass gap at lower orders:
\begin{eqnarray}
\Delta_1^{(4)} &=& \phantom{+}\frac{3(n-1)^4}{(2\pi)n^4}k_2 
\nonumber\\ & &
  +\frac{(n-1)^3}{(2\pi)^3n^4}\bigg\{h_1-2\bigg(
    \ln\frac{\pi}{\sqrt{2}}-\gamma\bigg)+3(2\pi)^2(k_1-k_2)\bigg\} 
\nonumber\\ & &
  +\frac{(n-1)^2}{(2\pi)^3n^4}\bigg\{-2h_1+h_2+4\bigg(
    \ln\frac{\pi}{\sqrt{2}}-\gamma\bigg)+\pi+3(2\pi)^2(k_0-k_1)\bigg\} 
\nonumber\\ & &
  +\frac{n-1}{(2\pi)^3n^4}\bigg\{h_1-h_2-2\bigg(
    \ln\frac{\pi}{\sqrt{2}}-\gamma\bigg)-\pi-3(2\pi)^2k_0\bigg\} \\
\Delta_2^{(4)} &=& -\frac{3(n-1)^4}{(2\pi)^3n^4}\bigg\{
    \ln\frac{\pi}{\sqrt{2}}-\gamma\bigg\} 
\nonumber\\ & &
  +\frac{(n-1)^3}{(2\pi)^3n^4}\bigg\{9\bigg(
    \ln\frac{\pi}{\sqrt{2}}-\gamma\bigg)+\frac12(3\pi+5)\bigg\} 
\nonumber\\ & &
  -\frac{(n-1)^2}{(2\pi)^3n^4}\bigg\{3\bigg(
    3\ln\frac{\pi}{\sqrt{2}}+\pi-3\gamma+1\bigg)+2\bigg\} 
\nonumber\\ & &
  +\frac{n-1}{(2\pi)^3n^4}\bigg\{1+\frac32\bigg(
    2\ln\frac{\pi}{\sqrt{2}}+\pi-2\gamma+1\bigg)\bigg\} \\
\Delta_3^{(4)} &=& \phantom{+}\frac{(n-1)^4}{(2\pi)^3n^4}
  -\frac{3(n-1)^3}{(2\pi)^3n^4}+\frac{3(n-1)^2}{(2\pi)^3n^4}
  -\frac{n-1}{(2\pi)^3n^4}
\end{eqnarray}
In order to determine the unknown coefficient $\Delta_0^{(4)}$,
we decompose it into $n$-independent constants:
\begin{equation}
\Delta_0^{(4)}=\frac{n-1}{n^4}\bigg\{s_0+(n-1)s_1+(n-1)^2s_2+(n-1)^3s_3\bigg\}.
\end{equation}
The coefficients $s_0,\cdots, s_3$ can then be determined by inserting 
$\Delta^{(4)}$ which was evaluated for different $L$'s
in eq.(\ref{condelta4}) and extrapolating to continuum limit.

The extrapolation of $s_k\,\,(k=0,\cdots,3)$ to $L=\infty$
can be done very efficiently with help of 
the procedure by L\"uscher and Weisz described in detail
in \cite{luescherextrap}.
We note that in our case
the constants $s_k$ have corrections of the form
\begin{equation}\label{korrtocont}
s_k^0(L) = s_k +
\sum_{p=1}^{\infty}\Big\{a_p+b_p\ln L+c_p(\ln L)^2+d_p(\ln L)^3\Big\}/L^{2p}.
\end{equation}
Through the extrapolation in the region $5\leq L\leq 20$, we obtain
\begin{eqnarray}\label{s0}
s_0 &=& \phantom{-}0.03954(1) \\
s_1 &=& \phantom{-}0.02903(1) \label{s1}\\
s_2 &=& \phantom{-}0.000756(1) \label{s2}\\
s_3 &=& -0.000649(1) \label{s3}
\end{eqnarray}

At this point, we would like to emphasize that 
for all four constants $s_k$ the limit $L\to\infty$ exists.
This is, of course, only the case if $\Delta^{(4)}$ calculated by us
contains the terms diverging logarithmically with $L$,
so that they cancel exactly 
with the other divergent factors of eq.(\ref{condelta4}) in a non-trivial way.
This correct logarithmic behavior of $\Delta^{(4)}$
is a strong consistency check on our evaluations of the diagrams.

\subsection{Conversion of the mass gap to the MS-scheme}

The $\beta$-function in the MS-scheme of dimensional regularization
is defined by
\begin{equation}\label{betams}
\beta(f) \equiv \mu\frac{\partial f}{\partial\mu}\bigg|_{f_0}
          = -f\sum_{\nu=1}^{\infty}b_{\nu}f^{\nu} ,
\end{equation}
where $f$ denotes the renormalized coupling and 
$\mu$ the normalization mass.
The first two coefficients in (\ref{betams}) are scheme independent:
\begin{eqnarray}
b_1 &=& \frac{n-2}{2\pi n}\\ 
b_2 &=& \frac{n-2}{(2\pi n)^2}
\end{eqnarray}
The remaining coefficients are known to 
four loops \cite{BrHiB,Hikami,BeWe,Weg}:
\begin{eqnarray}
b_3 &=& \frac{n-2}{(2\pi n)^3}\bigg[(n-2)\frac14+1\bigg] \\
b_4 &=& \frac{n-2}{(2\pi n)^4}\bigg[-(n-2)^2\frac{1}{12}+(n-2)u_1+u_2\bigg]
\label{b4ms}
\end{eqnarray}
with
\begin{eqnarray}
u_1 &=& \frac32(\upsilon+1)\\
u_2 &=& -\frac12(3\upsilon-1)
\end{eqnarray}
where the constant $\upsilon$ is given by $\upsilon\approx1.2020569$.

The corresponding $\Lambda$-parameter is
\begin{eqnarray}\label{lambdams}
\Lambda_{MS} &=& \mu(b_1f)^{-b_2/b_1^2}e^{-1/(b_1f)}\cdot\lambda(f) , \\
\lambda(f) &=& \exp\bigg[-\int_0^f dx
  \bigg(\frac{1}{\beta(x)}+\frac{1}{b_1x^2}-\frac{b_2}{b_1^2x}\bigg)\bigg] .
\end{eqnarray}
The function $\lambda(f)$ is well-defined at $f=0$ and can therefore
be expanded as a power series in $f$. Up to 4-loop, we find
\begin{eqnarray}
\Lambda_{MS} &=& \mu(b_1f)^{-b_2/b_1^2}e^{-1/(b_1f)}
  \bigg\{1+\frac{b_2^2-b_1b_3}{b_1^3}f \nonumber\\ & &
  +\frac{b_2^4-b_1^2b_2^3+2b_1^3b_2b_3-2b_1b_2^2b_3+b_1^2b_3^2-b_1^4b_4}
        {2b_1^6}f^2+{\cal O}(f^3)\bigg\}.
\end{eqnarray}

On the other hand, the $\beta$ function on the lattice is 
defined by\footnote{In this section, we introduce the lattice constant $a$.}
\begin{equation}\label{betalattice}
\hat{\beta}(f_0) \equiv -a\frac{\partial f_0}{\partial a}\bigg|_{f}
                 = -f_0\sum_{\nu=1}^{\infty}\hat{b}_{\nu}f_0^{\nu}.
\end{equation}
The coefficients here are also known to four loops \cite{FaTr,WBe,CaPe}:
\begin{eqnarray}
\hat{b}_1 &=& b_1\label{b1lattice}\\ 
\hat{b}_2 &=& b_2\label{b2lattice}\\
\hat{b}_3 &=& \frac{n-2}{(2\pi n)^3}[(n-2)h_1+h_2]\label{b3lattice}\\
\hat{b}_4 &=& \frac{n-2}{(2\pi n)^4}\Big[(n-2)^2t_1+(n-2)t_2+t_3\Big]
\label{b4lattice}
\end{eqnarray}
with
\begin{eqnarray}
h_1 &=& -0.088766484(1) \\
h_2 &=& 1+\pi/2-5\pi^2/24 \\
t_1 &=& -1.015(1) \\
t_2 &=& -5.44(1) \\
t_3 &=& -9.093756(4)
\end{eqnarray}
where the difficult 4-loop computation was performed recently by
Caracciolo and Pelissetto \cite{CaPe}.

The $\Lambda$-parameter on the lattice $\Lambda_L$ can be obtained 
from $\Lambda_{MS}$ of eq.(\ref{lambdams}) through the replacements
$\mu\to a^{-1}$, $f\to f_0$ and $\beta(x)\to\hat{\beta}(x)$. 
The two $\Lambda$-parameters are related by the formula:
\begin{equation}
\frac{\Lambda_L}{\Lambda_{MS}} =
\exp\frac12\bigg\{\ln\frac{\pi}{8}-\gamma-\frac{\pi}{n-2}\bigg\} .
\end{equation}

From this, it follows the relation between the
coupling constants in both schemes.
Up to order ${\cal O}(f^4)$, we find
\begin{eqnarray}
f_0 &=& f\bigg\{1+\sum_{\nu=1}^{\infty}X^{(\nu)}f^{\nu}\bigg\} \\
X^{(1)} &=& \frac{n-2}{4\pi n}\bigg[\ln\Big(\frac{\pi}{8}a^2\mu^2\Big)-
  \gamma\bigg]-\frac{1}{4n} \\
X^{(2)} &=& [X^{(1)}]^2 + \frac{1}{2\pi n}X^{(1)}
  + \frac{1}{(2\pi n)^2}\bigg[(n-2)\Big(h_1-\frac14\Big)+h_2-1\bigg]\\
X^{(3)} &=& [X^{(1)}]^3 + \frac{5}{4\pi n}[X^{(1)}]^2
  + \frac{1}{(2\pi n)^2}\bigg[(n-2)\Big(3h_1-\frac12\Big)+3h_2-2\bigg]X^{(1)}
\nonumber\\
& & +\frac{1}{(2\pi n)^3}\frac12
    \bigg[(n-2)^2\Big(t_1+\frac{1}{12}\Big)+(n-2)(t_2-u_1)+(t_3-u_2)\bigg]
\end{eqnarray}

Finally, we use this relation to express the mass gap 
[eqs. (\ref{condelta1})-(\ref{condelta3}) and (\ref{condelta4})]
in terms of the renormalized coupling in MS-scheme.
After a lengthy, but straightforward calculation we get
\begin{eqnarray}\label{mlexpan}
M(L) &=& \frac{n-1}{2nL}f
  \bigg\{1+\sum_{\nu=1}^{\infty}\kappa^{(\nu)}f^{\nu}\bigg\} \\
\kappa^{(1)} &=& \frac{n-2}{4\pi n}
               \bigg\{\ln\frac{\mu^2L^2}{4\pi}+\gamma\bigg\} \label{kp1}\\
\kappa^{(2)} &=& [\kappa^{(1)}]^2+\frac{1}{2\pi n}\kappa^{(1)}
  +3\frac{n-2}{(4\pi n)^2} \label{kp2}\\
\kappa^{(3)} &=& [\kappa^{(1)}]^3+\rho_1\bigg(\frac{n-2}{4\pi n}\bigg)^3
  +\frac{5}{4\pi n}[\kappa^{(1)}]^2
  +\rho_2\frac{2(n-2)}{(4\pi n)^2}\kappa^{(1)} \nonumber\\\label{kappa3}
      \label{kp3} & &
  +\rho_3\frac13\frac{(n-2)^2}{(4\pi n)^3}
  +\rho_4\frac{1}{(4\pi n)^2}\kappa^{(1)}
  +\rho_5\frac{n-2}{(4\pi n)^3}
  +\rho_6\frac{1}{(4\pi n)^3}
\end{eqnarray}
The expressions for the
coefficients $\rho_1,\cdots,\rho_6$ of $\kappa^{(3)}$ are rather long;
we therefore write them in appendix~B.
Here, the coefficients $\kappa^{(1)}$ and $\kappa^{(2)}$ are the result of
L\"uscher and Weisz \cite{LWU}.
By using dimensional regularization,
$\kappa^{(2)}$ was first calculated by Floratos and Petcher \cite{FlPe}
with less numerical precision.
Our result for $\kappa^{(3)}$ is the extension of their 1- and 
2-loop computations to 3-loop.

With dimensional regularization, $\kappa^{(\nu)}=0$ for $n=2$
since we formally have a free theory in this case.
$\kappa^{(1)}$ and $\kappa^{(2)}$ already satisfy this condition.
We checked that $\kappa^{(3)}$ is also zero within the errors,
which shows a further check on 
the correctness of our final results in the lattice calculations,
$s_0$, $s_1$, $s_2$ and $s_3$ [eqs.(\ref{s0})-(\ref{s3})].
This is, however, not only a check on our computations,
but also that on the 4-loop coefficient of the $\beta$ function 
on the lattice $\hat{b}_4$ 
which was introduced in our calculations by 
the conversion of the result on the lattice to the MS-scheme.

\subsection{Expansion in $z$}

In view of the evaluation of the finite volume mass gap $M(L)$
in the infinite volume limit, we introduce the variable
\begin{equation}\label{defz}
z = M(L)L
\end{equation}
and express the ratio $M(L)/\Lambda_{\overline{MS}}$ in terms of this
dimensionless and renormalization group invariant parameter
in perturbation theory for $z\to 0$, as suggested by L\"uscher \cite{Lup}.
Here, $\Lambda_{\overline{MS}}$ is related with $\Lambda_{MS}$ of
eq.(\ref{lambdams}) through
\begin{equation}\label{relmsmsbar}
\Lambda_{\overline{MS}} = \Lambda_{MS}
\exp\bigg[\frac12(\ln 4\pi-\gamma)\bigg] .
\end{equation}

For that purpose, we invert at first 
the $f$-expansion of eq.(\ref{mlexpan}) in the $z$-expansion:
\begin{equation}\label{fexptoz}
f = d_1z+d_2z^2+d_3z^3+d_4z^4+{\cal O}(z^5),
\end{equation}
where
\begin{eqnarray}
d_1 &=& \frac{2n}{n-1} \\
d_2 &=& -d_1^2\kappa^{(1)} \\
d_3 &=& -[2d_1d_2\kappa^{(1)}+d_1^3\kappa^{(2)}] \\
d_4 &=& -[(d_2^2+2d_1d_3)\kappa^{(1)}+3d_1^2d_2\kappa^{(2)}+d_1^4\kappa^{(3)}]
\end{eqnarray}
The insertion of this series in eq.(\ref{lambdams}), together with 
eq.(\ref{relmsmsbar}), yields finally
\begin{equation}\label{c0expgn}
\frac{M(L)}{\Lambda_{\overline{MS}}} =
\frac{e^{\gamma}}{4}\bigg(\frac{n-2}{n-1}\bigg)^{\frac{1}{n-2}}
  \bigg(\frac{z}{\pi}e^{\frac{\pi}{z}}\bigg)^{\frac{n-1}{n-2}}
  \bigg\{1+\sum_{\nu=1}^{\infty}a_{\nu}z^{\nu}\bigg\}
\end{equation}
with
\begin{eqnarray}
a_1 &=& \frac{1}{\pi(n-1)}\\\label{aal2}
a_2 &=& \frac{1}{24\pi^2(n-1)^2(n-2)}
(\alpha_1n^3 + \alpha_2n^2 + \alpha_3n + \alpha_4)
\end{eqnarray}
The expressions for the constants $\alpha_1,\cdots,\alpha_4$
are given in appendix~B.
We note that the coefficients $a_{\nu}$ in eq.(\ref{c0expgn})
are independent of $\mu$ and $L$
since $M(L)$, $\Lambda_{\overline{MS}}$ and $z$ are 
renormalization group invariants. In the above result,
the leading term and the first coefficient $a_1$
were calculated in ref. \cite{LWU},
and our determination of $a_2$ is the 3-loop correction to their
calculations.

\subsection{Definition of running coupling $\bar{g}^2(L)$}

Observing that the mass gap $M(L)$ in finite volume
[eq.(\ref{mlexpan})] is proportional to $f$ 
in the leading order, L\"uscher, Weisz and Wolff \cite{LuWeWo}
defined a coupling running with $L$ by
\begin{equation}\label{defrunningcoup}
\bar{g}^2(L) = 2nM(L)L/(n-1).
\end{equation}
We can then expand it,
for $\mu=1/L$, in the coupling of the MS-scheme:
\begin{equation}\label{runningcoup}
\bar{g}^2(L) = f + c_1f^2 + c_2f^3 + c_3f^4 + {\cal O}(f^5),
\end{equation}
where
\begin{eqnarray}
c_1 &=& -\frac{n-2}{4\pi n}[\ln(4\pi)-\gamma] \\
c_2 &=& c_1^2+\frac{c_1}{2\pi n}+3\frac{n-2}{(4\pi n)^2} \\
c_3 &=& c_1^3 + \rho_1\bigg(\frac{n-2}{4\pi n}\bigg)^3
  +\frac{5}{4\pi n}c_1^2 + \rho_2\frac{2(n-2)}{(4\pi n)^2}c_1 \nonumber\\ & &
  +\rho_3\frac13\frac{(n-2)^2}{(4\pi n)^3}
  +\rho_4\frac{1}{(4\pi n)^2}c_1
  +\rho_5\frac{n-2}{(4\pi n)^3}
\end{eqnarray}

For the coefficients of $\beta$-function
\begin{equation}
\tilde{\beta}(\bar{g}^2) \equiv -L\frac{\partial\bar{g}^2}{\partial L} 
  = -\bar{g}^2\sum_{l=1}^{\infty}\tilde{b}_l(\bar{g}^2)^l 
\end{equation}
we find
\begin{eqnarray}\label{tilb1}
\tilde{b}_1 &=& b_1 \\\label{tilb2}
\tilde{b}_2 &=& b_2 \\\label{tilb3}
\tilde{b}_3 &=& \frac{(n-1)(n-2)}{(2\pi n)^3} \\
\tilde{b}_4 &=& \frac{1}{4}\frac{n-2}{(2\pi n)^4}
\Big[(n-2)^3\chi_1 + (n-2)^2\chi_2 + (n-2)\chi_3 + \chi_4\Big] \label{tilb4}
\end{eqnarray}
where we have used the $\beta$-function in
MS-scheme [eqs.(\ref{betams})-(\ref{b4ms})].
The expressions for the constants $\chi_1,\cdots,\chi_4$ are listed
in appendix~B.

The corresponding $\Lambda$-parameter is given by
\begin{eqnarray} \label{lambdal}
\Lambda_{FV} &=& \frac{1}{L}(b_1\bar{g}^2)^{-b_2/b_1^2}e^{-1/(b_1\bar{g}^2)}
  \cdot\tilde{\lambda}(\bar{g}^2) , \\\label{lambllamb}
\tilde{\lambda}(\bar{g}^2) &=& \exp\bigg[-\int_0^{\bar{g}^2} dx
  \bigg(\frac{1}{\tilde{\beta}(x)}+\frac{1}{b_1x^2}
-\frac{b_2}{b_1^2x}\bigg)\bigg].
\end{eqnarray}
From the relation between two couplings in eq.(\ref{runningcoup}),
we can derive the ratio of $\Lambda$-parameters in both schemes:
\begin{equation}\label{lambdafv}
\frac{\Lambda_{FV}}{\Lambda_{MS}} = 
\frac{e^{\frac12\gamma}}{2\sqrt{\pi}} .
\end{equation}

\section{Results and Discussions}\label{cha3}
\setcounter{equation}{0}

In this chapter, we determine the
mass gap $m$ in units of the $\Lambda$-parameter 
by applying our results in the last chapter. 
For this, the following two approaches will be considered:
\begin{itemize}
\item 
Determination of the ratio $m/\Lambda_{\overline{MS}}$
by matching the behavior of the finite volume 
mass gap $M(L)$ evaluated at small $L$ with that of the mass shift
$\delta_0=[M(L)-m]/m$ known at large $L$
\item 
Determination of the ratio $\Lambda_{FV}/m$
by means of running coupling $\bar{g}^2(L)$
\end{itemize}
We discuss these two methods in detail and give the results.

After that, it is interesting to compare them with the analytic determination
performed by Hasenfratz, Maggiore and Niedermayer a few years ago
\cite{HaMaNi,HaNi}.
Their result in the general $\mbox{O}(n)$ model
has the following simple form:
\begin{equation}\label{onexachas}
m=\bigg(\frac{8}{e}\bigg)^{1/(n-2)}
\frac{1}{\Gamma[1+1/(n-2)]}\Lambda_{\overline{MS}} .
\end{equation}
In our work, we will need the values in the O(3) and O(4) models:
\begin{eqnarray}\label{o3exachas}
m &=& \bigg(\frac{8}{e}\bigg)\Lambda_{\overline{MS}} 
\hspace{1.2cm} \mbox{for} \hspace{0.5cm} \mbox{O}(3) \\\label{o4exachas}
m &=& \bigg(\frac{32}{\pi e}\bigg)^{1/2}\Lambda_{\overline{MS}}
\hspace{0.5cm} \mbox{for} \hspace{0.5cm} \mbox{O}(4)
\end{eqnarray}

By using the equations (\ref{lambdafv}) and (\ref{relmsmsbar}),
these formulas can be rewritten in terms of $\Lambda_{FV}$
defined in eq.(\ref{lambdal}) in connection with
the running coupling $\bar{g}^2(L)$:
\begin{eqnarray}
\frac{\Lambda_{FV}}{m} &=& \frac{e^{\gamma+1}}{32\pi}
\hspace{1.4cm} \mbox{for} \hspace{0.5cm} \mbox{O}(3) \\
\frac{\Lambda_{FV}}{m} &=& \frac{1}{16}\sqrt{\frac{e}{2\pi}}e^{\gamma}
\hspace{0.5cm} \mbox{for} \hspace{0.5cm} \mbox{O}(4)
\end{eqnarray}

\subsection{Determination of $m/\Lambda_{\overline{MS}}$ 
by applying $\delta_0$}

Our aim is to determine the mass gap $m$ in infinite volume 
by applying our result of the mass gap $M(L)$
in finite volume [eq.(\ref{c0expgn})].
For this purpose, we introduce two quantities $C_0(z)$ and $c_0$ by
\begin{eqnarray}\label{infvolm0}
M(L) &=& C_0(z)\Lambda_{\overline{MS}} \\\label{finvolm0}
m &=& c_0\Lambda_{\overline{MS}}
\end{eqnarray}
where $m$ is obtained from $M(L)$ by $m = \lim_{L\rightarrow\infty}M(L)$.

\begin{figure}[t]
\begin{center}
\leavevmode
\epsfxsize=91mm
\epsfbox{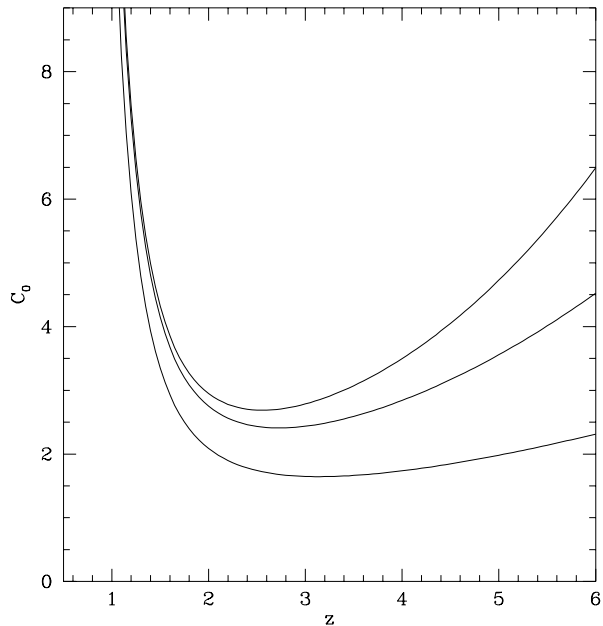}
\end{center}
\caption{$C_0$ as a function of $z$ in the O(3) model.
The lowest curve contains only the 1-loop contribution,
the middle one the contributions up to 2-loop and the uppermost curve up to
3-loop.
}
\label{lomt3}
\end{figure}

The problem is then the determination
of $c_0$ which may be done by attempting to extrapolate
$C_0(z)$ to the infinite volume limit, i.e. to $z\to\infty$.
Inspired by the successful application of the method
in the O($n$) model at $n=\infty$
where one can solve the given problem exactly,
L\"uscher \cite{Lup} tried, 
with his 1-loop calculation of $C_0(z)$, to estimate
$c_0$ in the O(3) model as the value of $C_0(z)$ 
at some intermediate region of $z$ (at around $3<z<4$)
with the hope that in this region the perturbative
$C_0(z)$ would be a good approximation for the infinite volume.
Here, we extend his 1-loop calculation to up to 3-loop
which is plotted in fig.~\ref{lomt3} as a function of $z$
in the O(3) model.
The lowest curve is the 1-loop result by L\"uscher while
the middle one comes from the 2-loop computation 
by Floratos and Petcher \cite{FlPe}.
Finally, the uppermost curve shows our 3-loop result.
The expected behavior of the curves, supported by the study in the 
O($\infty$) model, is that they rapidly decrease at small $z$ and 
then quickly become flat.
From the figure we, however, see that
it is difficult to make a clear estimation for $C_0(\infty)$ in this way.

For a better estimation of $c_0$, we make use of the mass shift defined by
\begin{equation} \label{delta0}
\delta_0 = \frac{M(L)-m}{m},
\end{equation}
as done in ref. \cite{FlPe}.
$\delta_0$ was first introduced by L\"uscher \cite{lueschercargese}
and can be calculated at large $\zeta=mL$ 
exactly up to an exponentially small correction term 
if the elastic forward scattering amplitude is known.\footnote{$\zeta$ is
related with $z$ by $z=(1+\delta_0)\zeta$ so that
$z\simeq\zeta$ at large $\zeta$.}
In the $\mbox{O}(n)$ $\sigma$-model,
the exact scattering matrix was determined
by brothers Zamolodchikov \cite{Za1,Za2}.
In particular, the forward scattering amplitude is known
and hence one can calculate $\delta_0$ in this model at large $\zeta$.

By using the definition of $\delta_0$ together with eq.(\ref{finvolm0}),
we obtain the following relation between $C_0(z)$ and $c_0$:
\begin{equation}
C_0 = (1+\delta_0)c_0 .
\end{equation}
From this, a formula expressing $c_0$ as a function of $\zeta$ follows:
\begin{equation}\label{c0zeta}
c_0(\zeta) = [1+\delta_0(\zeta)]^{-1}C_0[z(\zeta)].
\end{equation}
Actually, $c_0$ is independent of $\zeta$.
Therefore, it must be valid for all $\zeta$: 
$C_0[z(\zeta)]\sim 1+\delta_0(\zeta)$
if $C_0$ and $\delta_0$ are known exactly.
However, we know the two quantities only approximately;
$C_0$ for small $z$ and $\delta_0$, on the other hand, for large $\zeta$. 
Although in this approximate relation $c_0$ is in general
dependent on $\zeta$, there may exist the possibility that
one finds somewhere an overlapping region 
where both approximations are good and thus
$c_0$ becomes independent of $\zeta$.

\begin{figure}[htb]
\begin{center}
\leavevmode
\epsfxsize=90mm
\epsfbox{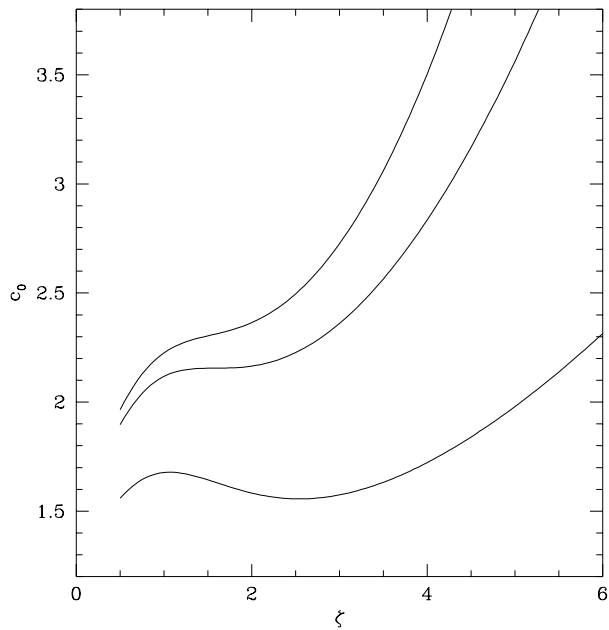}
\end{center}
\caption{$c_0$ as a function of $\zeta$ in the O(3) model.
The lowest curve contains for $C_0(z)$ only the 1-loop contribution,
the middle one the contributions up to 2-loop and the uppermost curve
up to 3-loop.
}
\label{zetan3inf}
\end{figure}

This idea works very well 
in the model for $n=\infty$, as shown in ref. \cite{FlPe}.
We therefore apply it to the determination of $c_0$ 
in the $\mbox{O}(n)$ model, 
explicitly in the O(3) and O(4) models.
In the O(3) model, $\delta_0$ was calculated 
by L\"uscher \cite{lueschercargese}:
\begin{equation}\label{delta0o3}
\delta_0 = 4\pi\int_{-\infty}^{\infty}
  dt\,e^{-\zeta\cosh t}\frac{\cosh t}{t^2+\frac94\pi^2}
  +{\cal O}(e^{-\kappa\zeta}),
\end{equation}
where $\kappa$ is a constant 
which is not smaller than $\sqrt{\frac32}$ and may become as large as 3.
If we now insert this quantity together with $C_0(z)$ at $n=3$ 
in eq.(\ref{c0zeta}), we obtain a functional relation of
$c_0$ in terms of $\zeta$ which
is illustrated in fig.~\ref{zetan3inf}.

The 1-loop result displays a region where $c_0\simeq 1.6$.
This value was the earlier estimation
by L\"uscher by this method \cite{Lup,lueschercargese}.
There, L\"uscher gave an optimistic error of 20~\%
which he apparently underestimated.
The 2-loop result corrects this by about 30~\% to $c_0\simeq 2.15$.
This is in agreement with the estimation by Floratos and Petcher
who give $c_0\simeq 2.1$ \cite{FlPe}.
Finally, we see from our 3-loop result
that the curve rises rather monotonically
instead of showing a wider flat area.
If this method should work well,
we would expect that the flat area at the 3-loop
becomes wider than at the 2-loop.
We attribute the bad applicability of this method 
to the fact that the convergence radius of $C_0(z)$ 
is apparently too small.
Nevertheless, we conservatively estimate
the 3-loop approximation to $c_0$ by $c_0\simeq 2.3$.
We note that it is very difficult to give a systematic error
on our estimation.

Our determination is to be compared with that by Hasenfratz et al.
which is $c_0 = 2.94304\cdots$ [see eq.(\ref{o3exachas})].
The value of $c_0\simeq 2.3$ at the 3-loop
still deviates noticeably from that by Hasenfratz et al..
One does not, however, need to be disturbed
because, from the convergence of the 1-, 2- and 3-loop approximations,
one sees a definite indication that 
the results in higher orders 
would arrive at that of Hasenfratz et al..

\begin{figure}[tbp]
\begin{center}
\leavevmode
\epsfxsize=90mm
\epsfbox{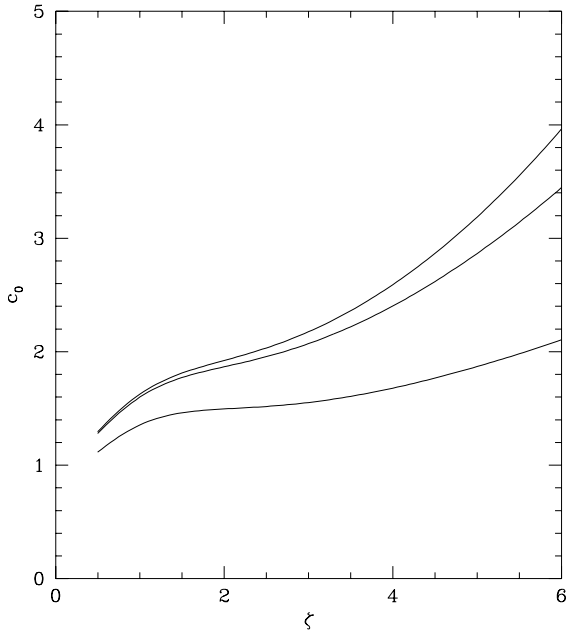}
\end{center}
\caption{$c_0$ as a function of $\zeta$ in the O(4) model.
The lowest curve contains for $C_0(z)$ only the 1-loop contribution,
the middle one the contributions up to 2-loop and the uppermost curve
up to 3-loop.
}
\label{zetan4inf}
\end{figure}

Now, we investigate the O(4) model and 
see how the picture changes in this model. 
The mass shift here has the form
\begin{equation}
\delta_0 = \frac{1}{2\pi}\int_{-\infty}^{\infty} dt
\cosh t\, e^{-\zeta\cosh t}
\bigg\{4-\bigg[\frac{2\pi t}{t^2+\frac{\pi^2}{4}}
\cdot\frac{B(\frac34-\frac{it}{2\pi},\frac34+\frac{it}{2\pi})}
     {B(\frac14-\frac{it}{2\pi},\frac14+\frac{it}{2\pi})}\bigg]^2\bigg\}
  +{\cal O}(e^{-\lambda\zeta}),
\end{equation}
where $B$ denotes the beta function and $\lambda$ is a constant whose value
lies in the same region as $\kappa$ in eq.(\ref{delta0o3}).
We again insert this quantity together with
$C_0(z)$ at $n=4$ in eq.(\ref{c0zeta}) to get the functional relation 
of $c_0$ in terms of $\zeta$. 
This relation is plotted in fig.~\ref{zetan4inf}.

The 1-loop result shows a stable region in the neighborhood $\zeta=3$,
from which one may estimate $c_0\simeq 1.5$.
If one, however, looks at the curves of the higher loops,
they become steeper and rise rather monotonically,
which makes it difficult to estimate the desired value.
We again see, as in the O(3) model, this method does not work very well.
However, a remarkable information is contained in the estimations of $c_0$
from the curves; by going to the higher loops, they 
are tending to converge to the value 
by Hasenfratz et al. of $c_0 = 1.93577\cdots$ [see eq.(\ref{o4exachas})].

From our determinations of the mass gap $m$, we conclude that 
in both O(3) and O(4) models our results support the correctness of
those by Hasenfratz et al..

\subsection{\mbox{Determination of $\Lambda_{FV}/m$} 
\mbox{by applying $\bar{g}^2(L)$}}

In chapter~\ref{cha2}, 
we introduced a coupling $\bar{g}^2(L)$ running with $L$. 
There, the coefficients of the corresponding
$\beta$-function $\tilde{\beta}(\bar{g}^2)$ were calculated
up to 4-loop order [eqs.(\ref{tilb1})-(\ref{tilb4})].
We make use of these perturbative coefficients to 
evaluate the $\Lambda$-parameter $\Lambda_{FV}$ in this scheme
in perturbation theory.

In order to obtain the perturbative expressions for $\Lambda_{FV}$,
we expand $\tilde{\lambda}(\bar{g}^2)$ 
of eq.(\ref{lambllamb}) in $\bar{g}^2$.
Up to order ${\cal O}(\bar{g}^4)$, we find
\begin{eqnarray}\label{2looplambdafv}
\Lambda^{(2)}_{FV} &=& 
  \frac{1}{L}(b_1\bar{g}^2)^{-b_2/b_1^2}e^{-1/(b_1\bar{g}^2)}\\
\Lambda^{(3)}_{FV} &=& 
  \frac{1}{L}(b_1\bar{g}^2)^{-b_2/b_1^2}e^{-1/(b_1\bar{g}^2)}
  \bigg\{1+\frac{b_2^2-b_1\tilde{b}_3}{b_1^3}\bar{g}^2\bigg\} \\
\label{4looplambdafv}
\Lambda^{(4)}_{FV} &=& 
  \frac{1}{L}(b_1\bar{g}^2)^{-b_2/b_1^2}e^{-1/(b_1\bar{g}^2)}
  \bigg\{1+\frac{b_2^2-b_1\tilde{b}_3}{b_1^3}\bar{g}^2 \nonumber\\ & &
  +\frac{b_2^4-b_1^2b_2^3+2b_1^3b_2\tilde{b}_3-2b_1b_2^2\tilde{b}_3
  +b_1^2\tilde{b}_3^2-b_1^4\tilde{b}_4}{2b_1^6}\bar{g}^4\bigg\}
\end{eqnarray}
where $\Lambda^{(2)}_{FV}$, $\Lambda^{(3)}_{FV}$ and $\Lambda^{(4)}_{FV}$
represent the 2-, 3- and 4-loop approximation to $\Lambda_{FV}$ respectively.
In the limit $L\to 0$ where
the coupling $\bar{g}^2(L)$ also goes to zero, they
converge to the exact
\begin{equation}
\Lambda_{FV} = \lim_{L\to 0}
\frac{1}{L}(b_1\bar{g}^2)^{-b_2/b_1^2}e^{-1/(b_1\bar{g}^2)}
\end{equation}
with the rates $\bar{g}^2(L)$, $\bar{g}^4(L)$ and $\bar{g}^6(L)$ respectively.

For determination of these perturbative $\Lambda$-parameters,
the values of the running coupling $\bar{g}^2(L)$ are needed.
In ref. \cite{LuWeWo}, they were measured numerically 
by L\"uscher, Weisz and Wolff through a finite size technique
for given $L$ in units of the mass gap $m$ in the O(3) model.
Furthermore, Hasenbusch made available to me
his new, more precise data \cite{Ha} which contain 
the measurements in the smaller region of the coupling.
Their results are listed in table~\ref{runningc}.
\begin{table}[tbp]\centering
\begin{tabular}{cc}\hline
\hspace{2cm} $mL$ \hspace{2cm} & \hspace{2cm} $\bar{g}^2(L)$ \hspace{2cm} 
\\ \hline
       0.0019(3)\phantom{3}            &   0.5372  \\
       0.0038(3)\phantom{3}            &   0.5747  \\
       0.0063(4)\phantom{3}            &   0.6060  \\
       0.0127(4)\phantom{3}            &   0.6553  \\
       0.0211(5)\phantom{3}            &   0.6970  \\
       0.0327(4)\phantom{3}            &   0.7383  \\
       0.0422(4)\phantom{3}            &   0.7646  \\
       0.0659(6)\phantom{3}            &   0.8166  \\
       0.1317(10)                      &   0.9176  \\
       0.2744(15)                      &   1.0595  \\       
       0.5557(13)                      &   1.2680  \\ \hline
\end{tabular}
\caption{\label{runningc} 
$\bar{g}^2(L)$ in the O(3) model as a function of $L$
in units of $m$}
\end{table}
In deriving the numerical values of the table,
the extrapolation to the continuum limit was done by fitting with 
a polynomial in $(a/L)^2$ which was supposed by Symanzik \cite{Sy}.

\begin{figure}[p]
\begin{center}
\leavevmode
\epsfxsize=112mm
\epsfbox{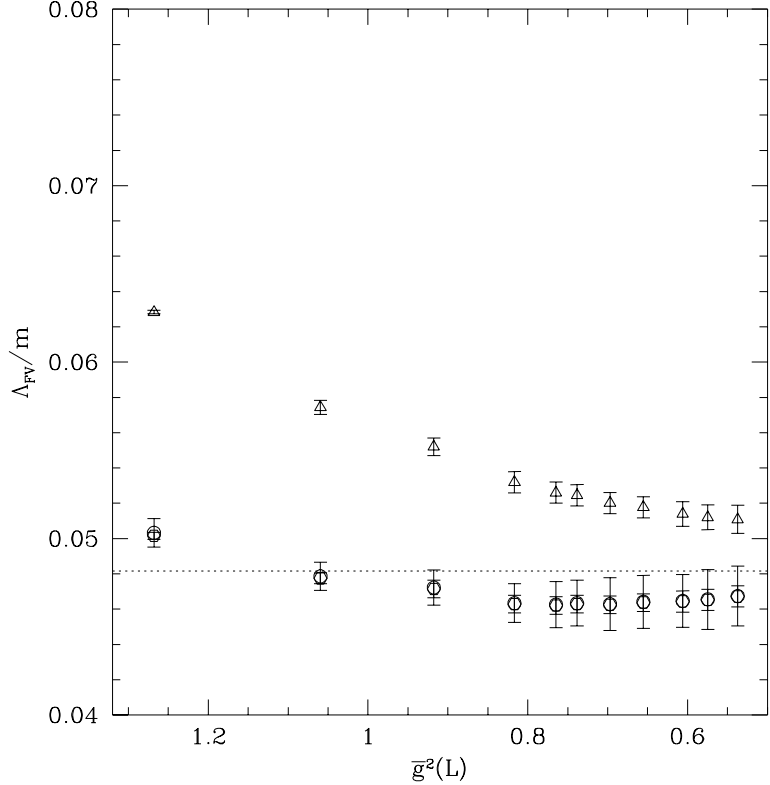}
\end{center}
\caption{
2-, 3- and 4-loop approximations to $\Lambda_{FV}/m$ in the O(3) model
from table~\ref{lambapp}.
The triangular points show the 2-loop approximation
to the $\Lambda$-parameter, the hexagonal points the 3-loop
and the circular points the 4-loop approximation.
The dotted line is the value of Hasenfratz et al. 
($\Lambda_{FV}/m=0.04815\cdots$).
}
\label{fvml}
\end{figure}

If we insert the numbers of table~\ref{runningc}
in eqs.(\ref{2looplambdafv})-(\ref{4looplambdafv}),
we get $\Lambda_{FV}/m$ in 2-, 3- and 4-loop approximations.
The results are listed in table~\ref{lambapp} and also illustrated in 
fig.~\ref{fvml}. 
\begin{table}[p]\centering
\begin{tabular}{c|ccc}\hline
\hspace{0.5cm} \mbox{} $\bar{g}^2(L)$ \mbox{} \hspace{0.5cm} & 
\hspace{0.5cm} \mbox{} $\Lambda^{(2)}_{FV}/m$ \mbox{} \hspace{0.5cm} &  
\hspace{0.5cm} \mbox{} $\Lambda^{(3)}_{FV}/m$ \mbox{} \hspace{0.5cm} &  
\hspace{0.5cm} \mbox{} $\Lambda^{(4)}_{FV}/m$ \mbox{} \hspace{0.5cm} 
\\ \hline
 0.5372  &  0.0511(8)  &   0.0467(6)  &   0.0467(17)  \\
 0.5747  &  0.0512(7)  &   0.0465(6)  &   0.0466(17)  \\
 0.6060  &  0.0514(7)  &   0.0464(6)  &   0.0465(15)  \\
 0.6553  &  0.0518(6)  &   0.0464(5)  &   0.0464(15)  \\
 0.6970  &  0.0520(6)  &   0.0462(5)  &   0.0463(15)  \\
 0.7383  &  0.0525(6)  &   0.0463(5)  &   0.0463(13)  \\
 0.7646  &  0.0526(6)  &   0.0462(5)  &   0.0463(13)  \\
 0.8166  &  0.0532(6)  &   0.0463(5)  &   0.0463(11)  \\
 0.9176  &  0.0552(5)  &   0.0471(5)  &   0.0472(10)  \\
 1.0595  &  0.0574(4)  &   0.0478(3)  &   0.0479(8)\phantom{0}  \\
 1.2680  &  0.0628(1)  &   0.0502(2)  &   0.0503(8)\phantom{0}
\\ \hline
\end{tabular}
\caption{\label{lambapp}
2-, 3- and 4-loop approximations to $\Lambda_{FV}/m$
in the O(3) model}
\end{table}
In the figure, we also draw the analytic value of Hasenfratz et al. 
for comparison.
We see in the figure the difference between 
the 3- and 4-loop results is very small.
Nevertheless, our 4-loop result lies between the
2- and 3-loop ones, so that by consideration of the higher
order for $\Lambda_{FV}$ the ratio $\Lambda_{FV}/m$ 
lies nearer to the analytic value.
The remarkable fact, however, is that 
the 3- and 4-loop curves which, with the old data by L\"uscher et al. alone,
seemed to fall down continuously 
again rise together to the compared value if one adds the new data 
for smaller coupling 
by Hasenbusch beginning from $\bar{g}^2(L)=0.6970$.
We can thus expect very well that in the limit
$\bar{g}^2(L)\to 0$ the perturbative $\Lambda_{FV}/m$ 
would converge to the result of Hasenfratz et al..

To give a further impression on the systematic errors, we consider 
an alternative definition of $n$-loop approximations to $\Lambda_{FV}$;
instead of expanding $\tilde{\lambda}(\bar{g}^2)$ in $\bar{g}^2$,
we insert the perturbative coefficients of 
$\tilde{\beta}(x)$ in $\tilde{\lambda}(\bar{g}^2)$ and integrate this exactly.
If we apply the coefficients calculated up to 4-loop order, 
we obtain the following approximations to $\Lambda_{FV}$ at
2-, 3- and 4-loop levels
which are a little modified from
eqs.(\ref{2looplambdafv})-(\ref{4looplambdafv}):
\begin{eqnarray} \label{lambdam}
\bar{\Lambda}^{(2)}_{FV} &=& 
  \frac{1}{L}(b_1\bar{g}^2)^{-b_2/b_1^2}e^{-1/(b_1\bar{g}^2)} \nonumber\\ & &
  \cdot\exp\bigg[-\int_0^{\bar{g}^2}dx
  \bigg(\frac{1}{x^2(b_1+b_2x)}+\frac{1}{b_1x^2}-\frac{b_2}{b_1^2x}\bigg)\bigg]
\\
\bar{\Lambda}^{(3)}_{FV} &=& 
  \frac{1}{L}(b_1\bar{g}^2)^{-b_2/b_1^2}e^{-1/(b_1\bar{g}^2)} \nonumber\\ & &
  \cdot\exp\bigg[-\int_0^{\bar{g}^2}dx
  \bigg(\frac{1}{x^2(b_1+b_2x+\tilde{b}_3x^2)}
  +\frac{1}{b_1x^2}-\frac{b_2}{b_1^2x}\bigg)\bigg] \\
\bar{\Lambda}^{(4)}_{FV} &=& 
  \frac{1}{L}(b_1\bar{g}^2)^{-b_2/b_1^2}e^{-1/(b_1\bar{g}^2)} 
\nonumber\\\label{lambdam4} & &
  \cdot\exp\bigg[-\int_0^{\bar{g}^2}dx
  \bigg(\frac{1}{x^2(b_1+b_2x+\tilde{b}_3x^2+\tilde{b}_4x^3)}
  +\frac{1}{b_1x^2}-\frac{b_2}{b_1^2x}\bigg)\bigg] 
\end{eqnarray}

\begin{figure}[tbp]
\begin{center}
\leavevmode
\epsfxsize=122mm
\epsfbox{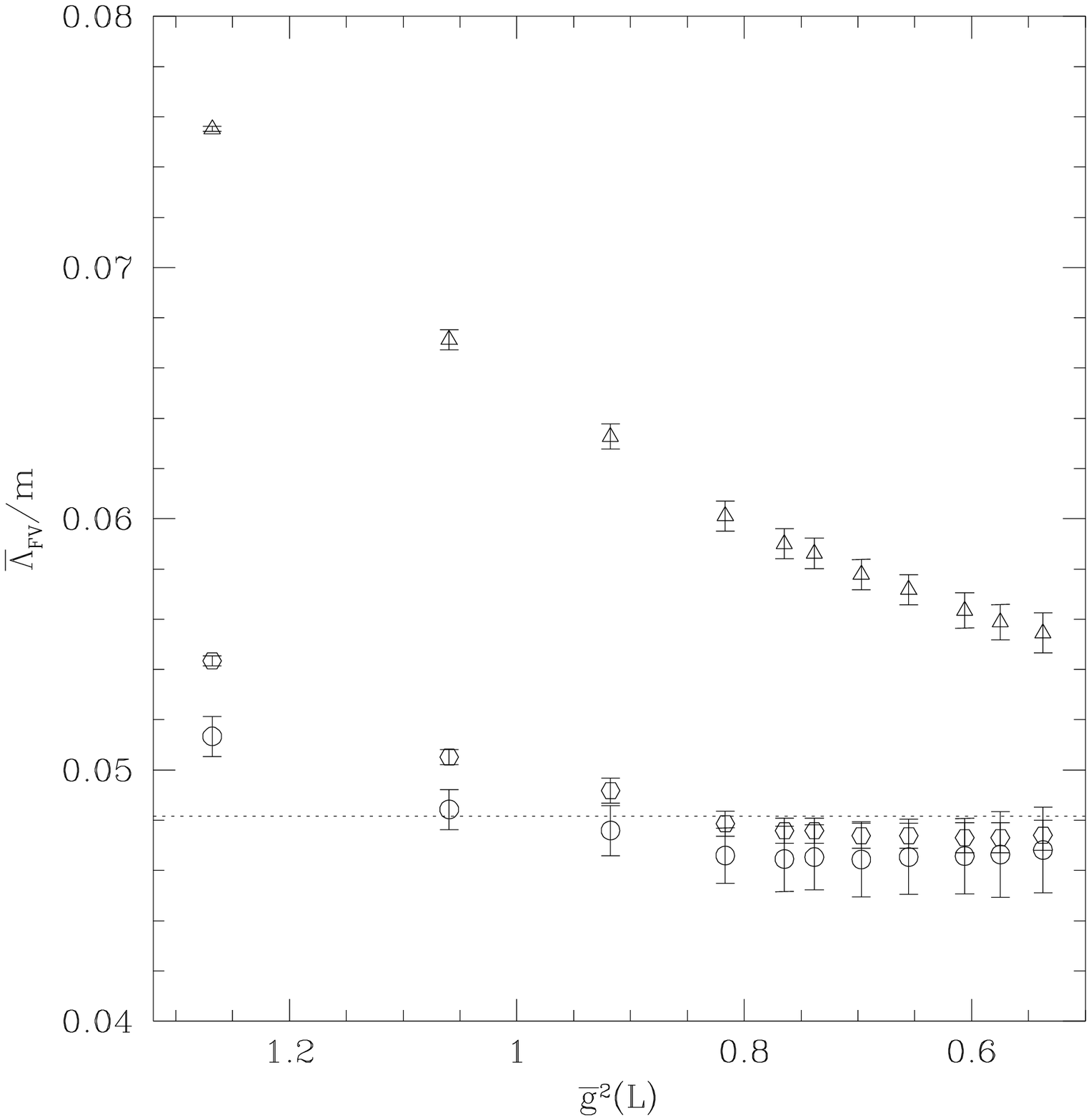}
\end{center}
\caption{
2-, 3- and 4-loop approximations to $\Lambda_{FV}/m$ in the O(3) model
from table~\ref{lambapm}.
The arrangement of the points and dotted line is same as 
in fig.~\ref{fvml}.
}
\label{fvmlmo}
\end{figure}

Insertion of the data of table~\ref{runningc} in these equations
gives the numbers in table~\ref{lambapm} which
are also plotted in fig.~\ref{fvmlmo}.
\begin{table}[p]\centering
\begin{tabular}{c|ccc}\hline
\hspace{0.5cm} \mbox{} $\bar{g}^2(L)$ \mbox{} \hspace{0.5cm} &
\hspace{0.5cm} \mbox{} $\bar{\Lambda}^{(2)}_{FV}/m$ \mbox{} \hspace{0.5cm} &
\hspace{0.5cm} \mbox{} $\bar{\Lambda}^{(3)}_{FV}/m$ \mbox{} \hspace{0.5cm} &
\hspace{0.5cm} \mbox{} $\bar{\Lambda}^{(4)}_{FV}/m$ \mbox{} \hspace{0.5cm} 
\\ \hline
 0.5372  & 0.0555(8) &  0.0474(6)  &  0.0468(17)  \\
 0.5747  & 0.0559(7) &  0.0473(6)  &  0.0466(17)  \\
 0.6060  & 0.0563(7) &  0.0473(6)  &  0.0466(15)  \\
 0.6553  & 0.0572(6) &  0.0474(5)  &  0.0465(15)  \\
 0.6970  & 0.0578(6) &  0.0474(5)  &  0.0464(15)  \\
 0.7383  & 0.0586(6) &  0.0476(5)  &  0.0465(23)  \\
 0.7646  & 0.0590(6) &  0.0476(5)  &  0.0465(13)  \\
 0.8166  & 0.0601(6) &  0.0479(5)  &  0.0466(11)  \\
 0.9176  & 0.0633(5) &  0.0492(5)  &  0.0476(10)  \\
 1.0595  & 0.0671(4) &  0.0505(3)  &  0.0484(8)\phantom{0}  \\
 1.2680  & 0.0755(1) &  0.0544(2)  &  0.0513(8)\phantom{0}
\\ \hline
\end{tabular}
\caption{\label{lambapm}
Alternative 2-, 3- and 4-loop approximations to $\Lambda_{FV}/m$
in the O(3) model}
\end{table}
Apart from the fact that the difference between the 3- and 4-loop results is
a little larger, we have here almost the same picture as in fig.~\ref{fvml}.
Although the 4-loop result lies below the 3-loop one
so that it deviates a little more from the value of Hasenftratz et al.,
there is no cause for alarm
since both curves approach the analytic value together. 
From our study through the alternative definition of the perturbative 
$\Lambda_{FV}$, we therefore conclude that 
our determination of $\Lambda_{FV}/m$, as in the first case, 
agrees with that by Hasenfratz et al. very well.

\begin{figure}[p]
\begin{center}
\leavevmode
\epsfxsize=122mm
\epsfbox{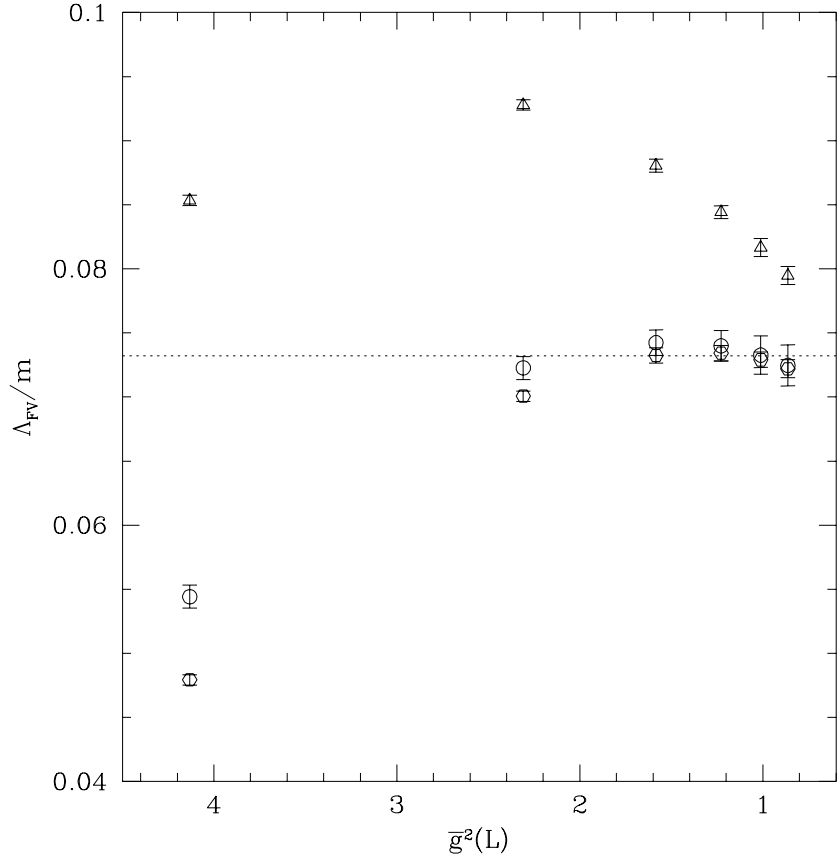}
\end{center}
\caption{
2-, 3- and 4-loop approximations to $\Lambda_{FV}/m$ in the O(4) model
from table~\ref{lambappo4f}.
The arrangement of the points is same as in fig.~\ref{fvml}.
The dotted line is the value of Hasenfratz et al. 
($\Lambda_{FV}/m=0.07321\cdots$).
}
\label{fvmlo4}
\end{figure}

In ref. \cite{HaHo}, Hasenbusch and Horgan measured the 
running coupling $\bar{g}^2(L)$ as a function of $L$ in units of 
the mass gap $m$ in the O(4) model by using the finite-size scaling analysis
of L\"uscher et al. \cite{LuWeWo}. 
We make use of the measurements
to determine the ratio $\Lambda_{FV}/m$ in this model.
Their data for the running coupling are given in table~\ref{runningco4}.
\renewcommand{\arraystretch}{1.2}
\begin{table}[t]\centering
\begin{tabular}{cc}\hline
\hspace{2cm} $mL$ \hspace{2cm} & \hspace{2cm} $\bar{g}^2(L)$ \hspace{2cm} 
\\ \hline
       1/8            &   0.863(2)\phantom{0} \\
       1/4            &   1.011(2)\phantom{0} \\
       1/2            &   1.228(2)\phantom{0} \\
       1              &   1.584(4)\phantom{0} \\
       2              &   2.309(10)  \\
       4              &   4.132(10)  \\ \hline
\end{tabular}
\caption{\label{runningco4} 
$\bar{g}^2(L)$ in the O(4) model as a function of $L$
in units of $m$}
\end{table}
If we apply the numbers in the table
to the eqs.(\ref{2looplambdafv})-(\ref{4looplambdafv}),
we obtain $\Lambda_{FV}/m$ in 2-, 3- and 4-loop approximations.
They are written in table~\ref{lambappo4f}
and also illustrated in fig.~\ref{fvmlo4}.
\renewcommand{\arraystretch}{1.2}
\begin{table}[p]\centering
\begin{tabular}{c|ccc}\hline
\hspace{0.5cm} \mbox{} $\bar{g}^2(L)$ \mbox{} \hspace{0.5cm} &
\hspace{0.5cm} \mbox{} $\Lambda^{(2)}_{FV}/m$ \mbox{} \hspace{0.5cm} &
\hspace{0.5cm} \mbox{} $\Lambda^{(3)}_{FV}/m$ \mbox{} \hspace{0.5cm} &
\hspace{0.5cm} \mbox{} $\Lambda^{(4)}_{FV}/m$ \mbox{} \hspace{0.5cm} 
\\ \hline
 0.863  & 0.0795(7)  &    0.0722(7) &   0.0725(16) \\ 
 1.011  & 0.0817(7)  &    0.0729(6) &   0.0733(15) \\ 
 1.228  & 0.0844(5)  &    0.0734(6) &   0.0740(12) \\
 1.584  & 0.0880(5)  &    0.0733(6) &   0.0742(10) \\
 2.309  & 0.0928(4)  &    0.0701(4) &   0.0723(9)\phantom{0} \\
 4.132  & 0.0853(4)  &    0.0479(4) &   0.0544(9)\phantom{0} 
\\ \hline
\end{tabular}
\caption{\label{lambappo4f}
2-, 3- and 4-loop approximations to $\Lambda_{FV}/m$
in the O(4) model}
\end{table}
The figure shows very good agreement between 
our result and the one by Hasenfratz et al..
Already at $\bar{g}^2(L)=1.584$, $\Lambda_{FV}/m$ at 4-loop level
is almost the same as the analytic value.

\begin{figure}[tbp]
\begin{center}
\leavevmode
\epsfxsize=122mm
\epsfbox{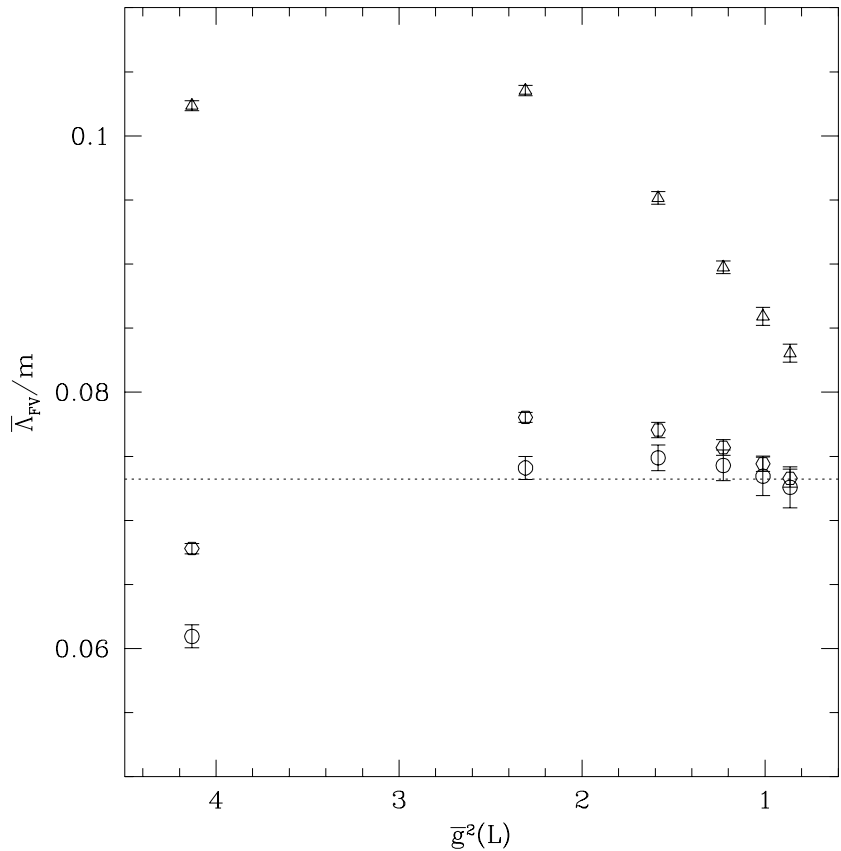}
\end{center}
\caption{
2-, 3- and 4-loop approximations to $\Lambda_{FV}/m$ in the O(4) model
from table~\ref{lambappo4s}.
The arrangement of the points and dotted line is same as 
in fig.~\ref{fvmlo4}.
}
\label{fvmlo4m}
\end{figure}

As in the O(3) model, by inserting the data of table~\ref{runningco4}
in the eqs.(\ref{lambdam})-(\ref{lambdam4}),
we also determine $\Lambda_{FV}/m$ by means of the modified definition
of the perturbative $\Lambda_{FV}$.
The results are written in table~\ref{lambappo4s} 
and also plotted in fig.~\ref{fvmlo4m}.
Apart from the fact that the curves for 
the 3- and 4-loop approximations in the figure
changed their places,
we see here the same picture as in fig.~\ref{fvmlo4}.
The agreement of two values at a little smaller 
$\bar{g}^2(L)$ is irrelevant from the fact that
we are at the end interested in the determination of $\Lambda_{FV}/m$ at
the limit $\bar{g}^2(L)\to 0$.

As a whole, we conclude that
not only in the O(3) but also in the O(4) model 
our determinations of $\Lambda_{FV}/m$ agree 
with those by Hasenfratz et al. very well.

\section*{Acknowledgement}

I thank Peter Weisz for suggesting this work and many useful discussions.
He also read through this manuscript, which is greatly acknowledged as well.
I express my thanks also to Martin Hasenbusch for having made available to me 
his unpublished numerical data.

\vspace{4cm}

\renewcommand{\arraystretch}{1.2}
\begin{table}[tbp]\centering
\begin{tabular}{c|ccc}\hline
\hspace{0.5cm} \mbox{} $\bar{g}^2(L)$ \mbox{} \hspace{0.5cm} &
\hspace{0.5cm} \mbox{} $\bar{\Lambda}^{(2)}_{FV}/m$ \mbox{} \hspace{0.5cm} &
\hspace{0.5cm} \mbox{} $\bar{\Lambda}^{(3)}_{FV}/m$ \mbox{} \hspace{0.5cm} &
\hspace{0.5cm} \mbox{} $\bar{\Lambda}^{(4)}_{FV}/m$ \mbox{} \hspace{0.5cm} 
\\ \hline
 0.863  & 0.0830(7) &   0.0733(7) &   0.0726(16) \\
 1.011  & 0.0859(7) &   0.0744(6) &   0.0734(15) \\
 1.228  & 0.0898(5) &   0.0757(6) &   0.0743(12) \\
 1.584  & 0.0952(5) &   0.0771(6) &   0.0749(10) \\
 2.309  & 0.1035(4) &   0.0780(4) &   0.0741(9)\phantom{0}\\
 4.132  & 0.1024(4) &   0.0678(4) &   0.0609(9)\phantom{0} 
\\ \hline
\end{tabular}
\caption{\label{lambappo4s}
Alternative 2-, 3- and 4-loop approximations to $\Lambda_{FV}/m$
in the O(4) model}
\end{table}

\newpage

\begin{appendix}

\section{Feynman diagrams}

We list here
the 2- and 3-loop Feynman diagrams contributing to the mass gap 
in fourth order.
In the diagrams, the internal dot denotes the vertices while the cross
means those coming from ``$\pi^2$ insertion.''

\begin{figure}[h]
\begin{center}
\leavevmode
\epsfxsize=96mm
\epsfbox{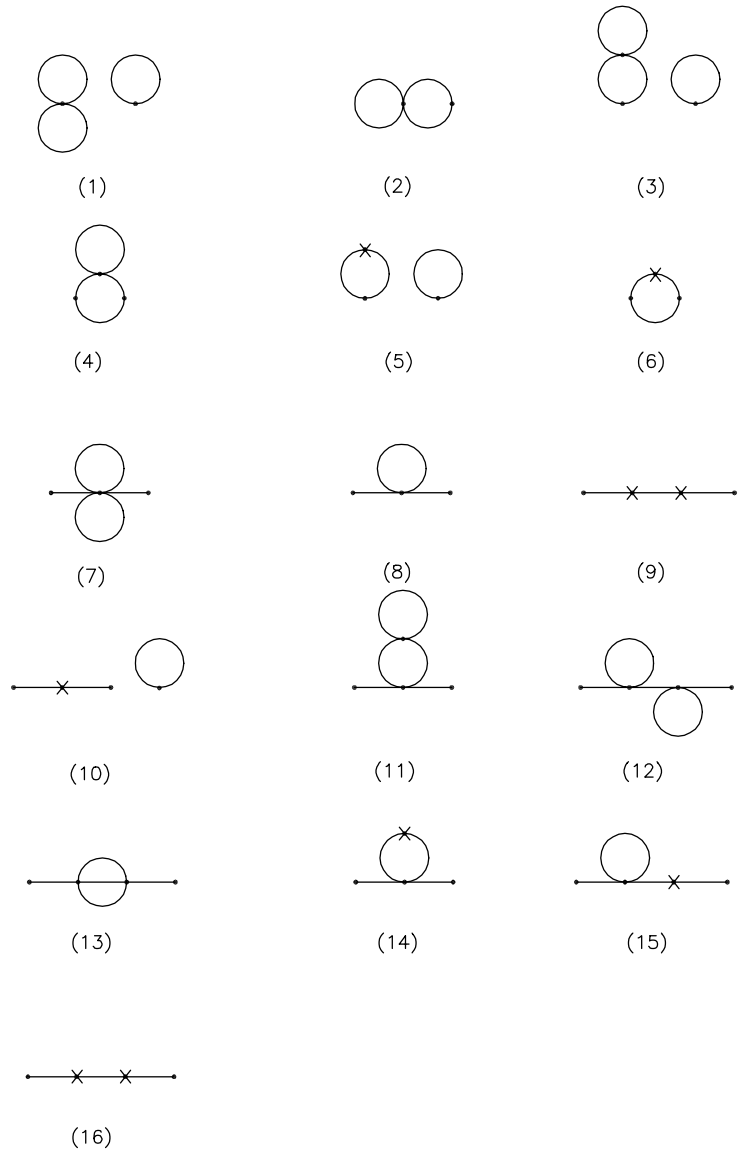}
\end{center}
\caption{2-loop diagrams}
\label{fd2}
\end{figure}

\begin{figure}[p]
\begin{center}
\leavevmode
\epsfxsize=130mm
\epsfbox{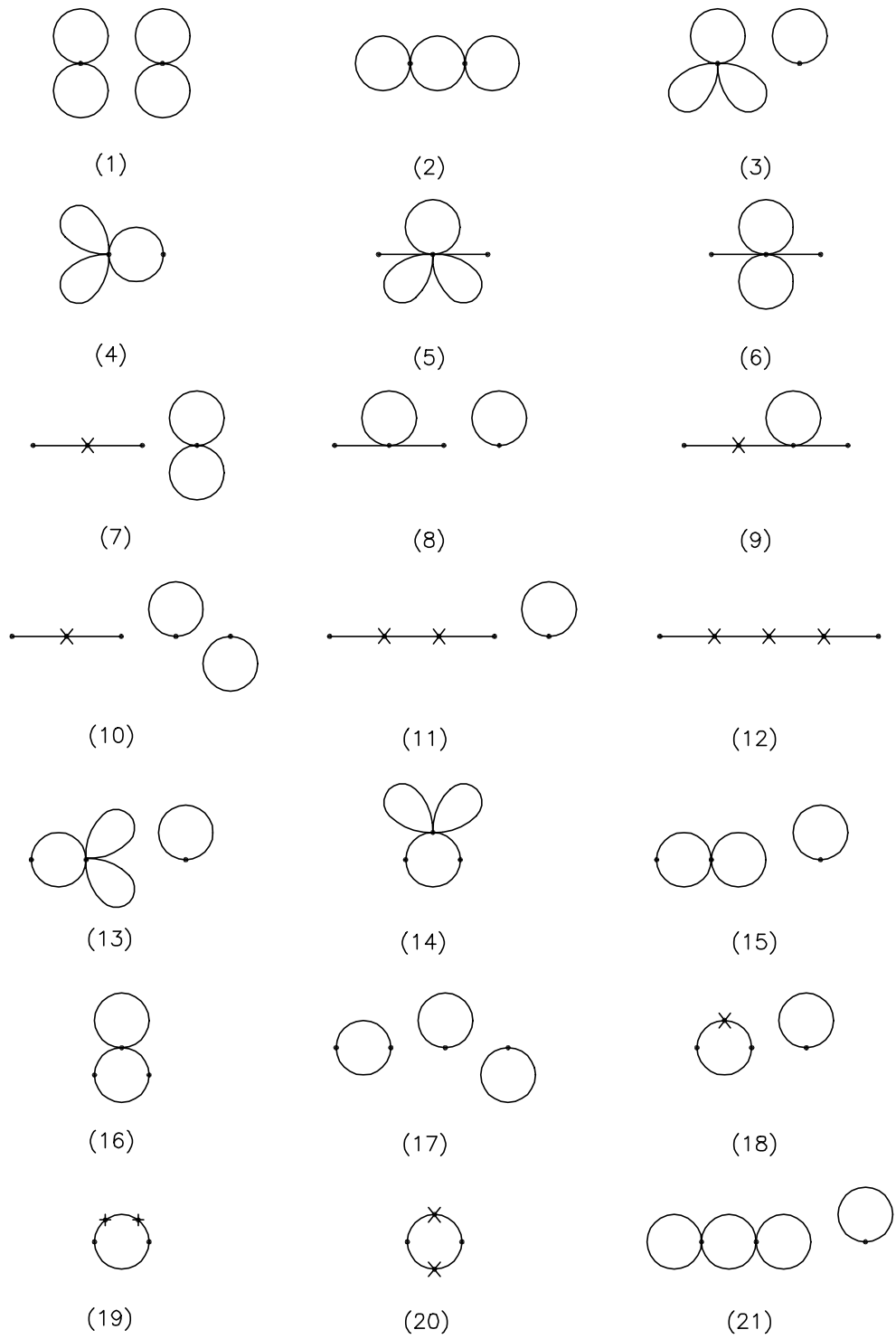}
\end{center}
\caption{3-loop diagrams}
\label{fd3}
\end{figure}

\setcounter{figure}{8}

\begin{figure}[p]
\begin{center}
\leavevmode
\epsfxsize=114mm
\epsfbox{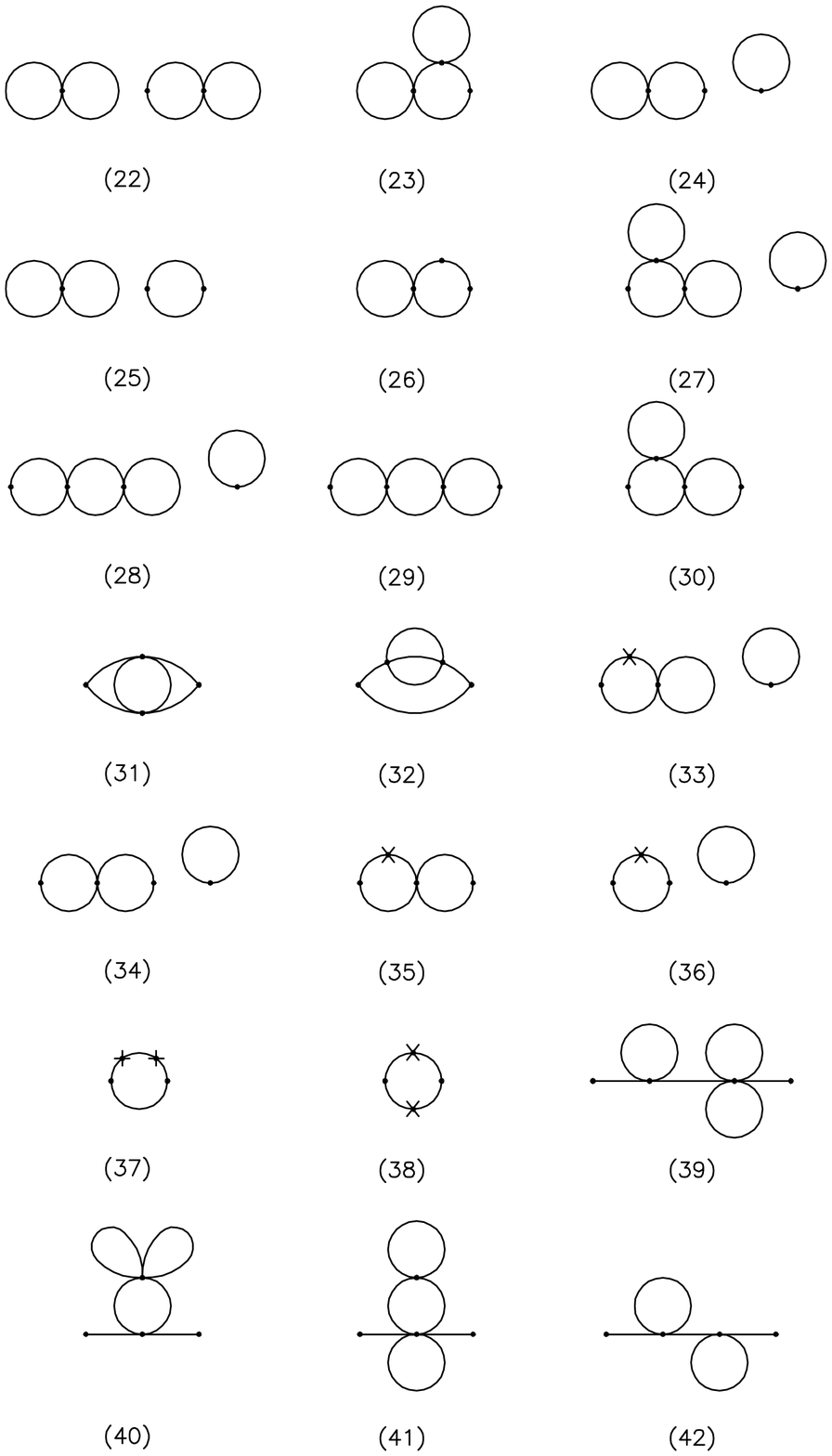}
\end{center}
\caption{3-loop diagrams (continued)}
\end{figure}

\setcounter{figure}{8}

\begin{figure}[p]
\begin{center}
\leavevmode
\epsfxsize=120mm
\epsfbox{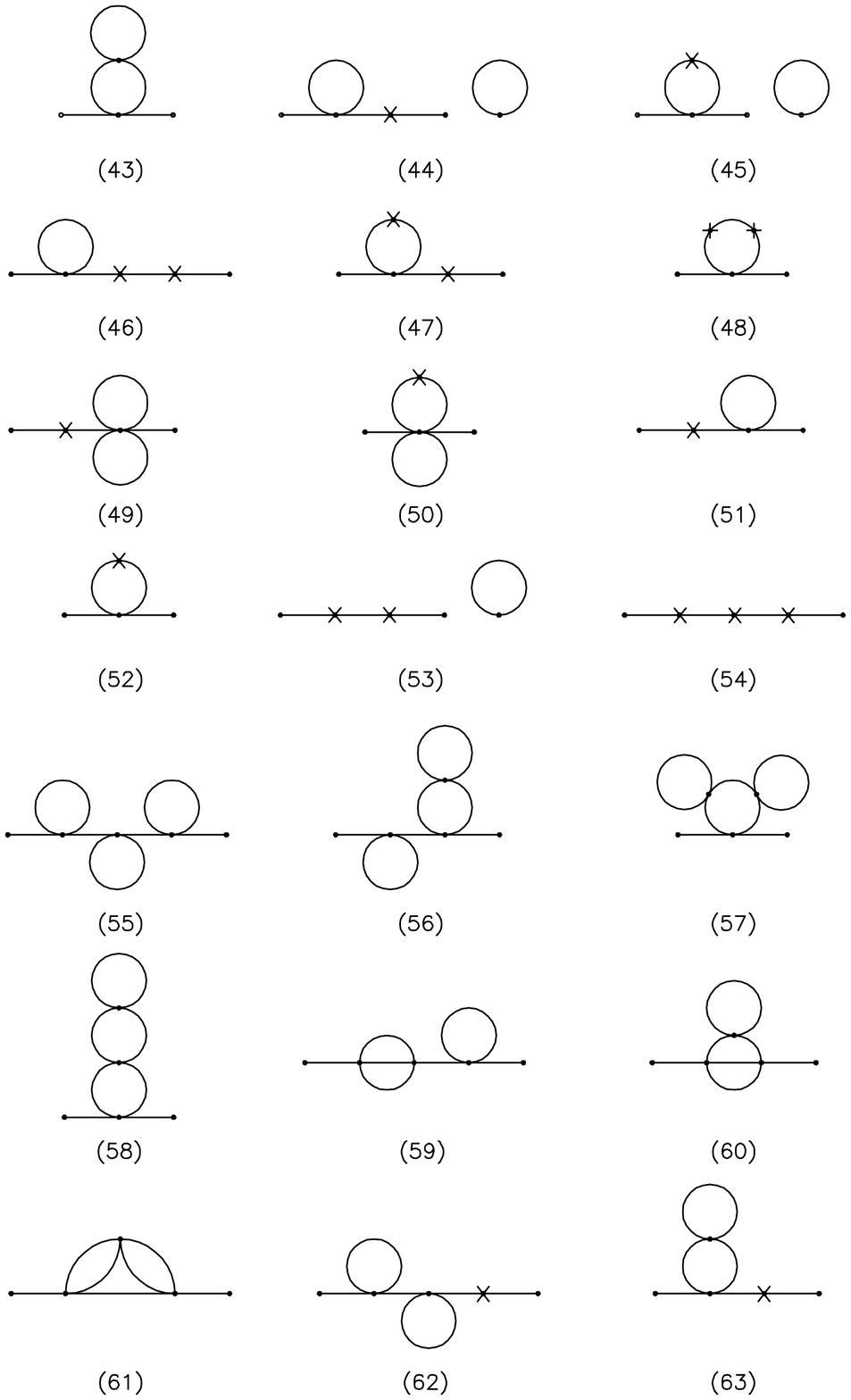}
\end{center}
\caption{3-loop diagrams (continued)}
\end{figure}

\setcounter{figure}{8}

\begin{figure}[t]
\begin{center}
\leavevmode
\epsfxsize=120mm
\epsfbox{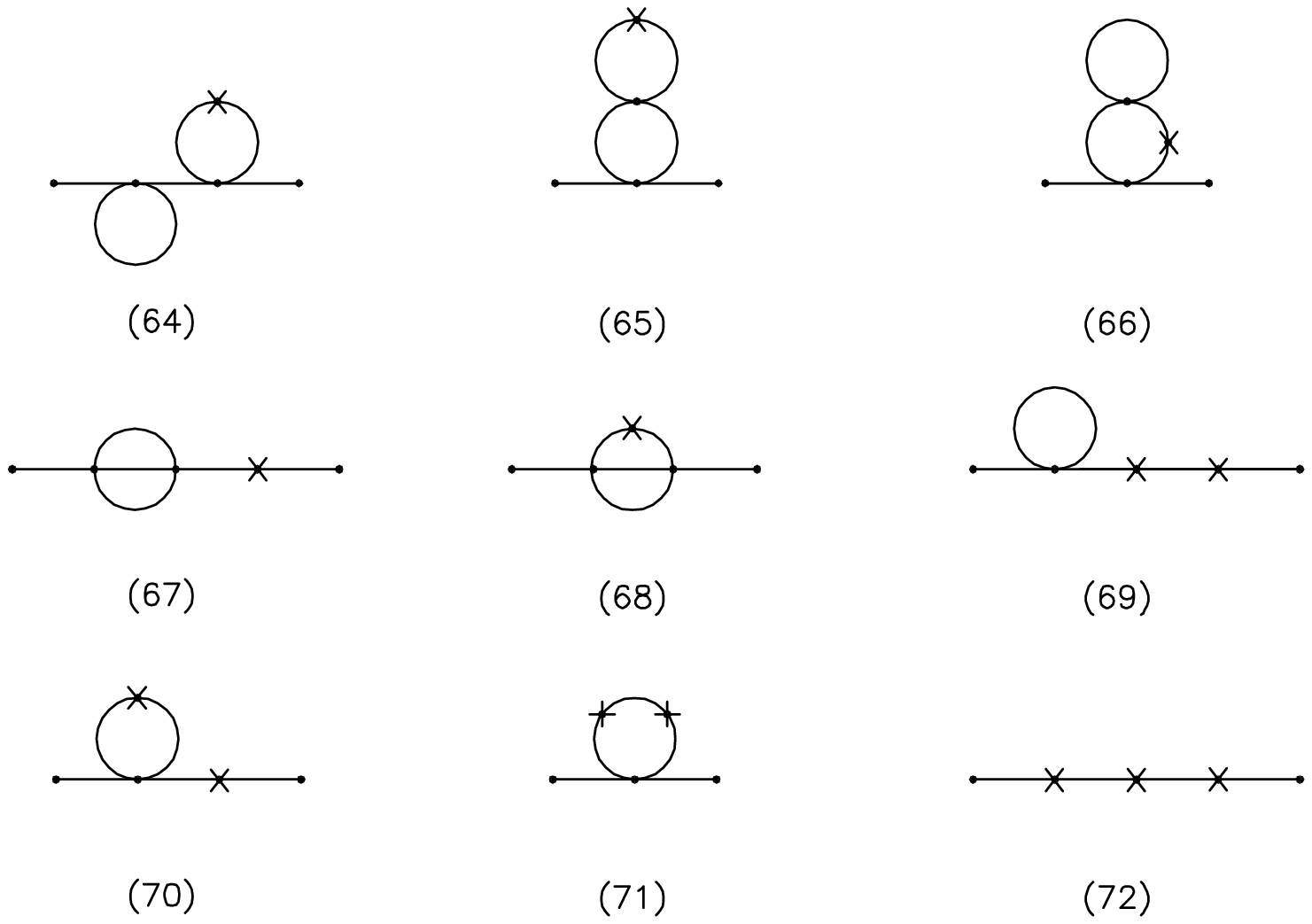}
\end{center}
\caption{3-loop diagrams (continued)}
\end{figure}

\newpage
\section{Expressions of Coefficients}

In this appendix, we show the expressions of the coefficients 
defined in the main text.

\subsection{Expressions of the coefficients appearing in eq.(\ref{kappa3})}

\begin{eqnarray*}
\rho_1 &=& \bigg(\ln\frac{\pi^2}{2}-2\gamma\bigg)^3+(4\pi)^3s_3 \\
\rho_2 &=& 3\bigg(\ln\frac{\pi^2}{2}\bigg)^2 
+3(\pi-4\gamma+1)\ln\frac{\pi^2}{2} 
- 6\gamma(\pi+1) \nonumber\\
& & +24k_1\pi^2 + 6h_1 + 12\gamma^2 -1 \\
\rho_3 &=& 
 18\bigg(\ln\frac{\pi^2}{2}\bigg)^3  
 + 3(3\pi- 36\gamma+1)\bigg(\ln\frac{\pi^2}{2}\bigg)^2 \nonumber\\
&& +12\Big[12k_1\pi^2+h_1+18\gamma^2-\gamma(3\pi+1)\Big]\ln\frac{\pi^2}{2}
\nonumber\\
& & +192\pi^3(s_2 + 3s_3)-144\gamma^3 + 12\gamma^2(3\pi+1) \nonumber\\
& & -24\gamma(12\pi^2k_1+h_1)+12t_1+1 \\
\rho_4 &=& 3\bigg(\ln\frac{\pi^2}{2}-2\gamma\bigg)^2
+ 3(16k_0 + 16k_1-1)\pi^2 + 12h_2 - 6\pi -8 \\
\rho_5 &=& 
  3\bigg(\ln\frac{\pi^2}{2}\bigg)^3 
 -6(3\gamma+\pi)\bigg(\ln\frac{\pi^2}{2}\bigg)^2 \nonumber\\
& & +2\Big[3(8k_0 + 8k_1-1)\pi^2+(12\gamma-1)\pi+2h_2+18\gamma^2\Big]
\ln\frac{\pi^2}{2} \nonumber\\ 
& & +16\pi^3(4s_1+8s_2+12s_3-3k_1) -
    12\gamma\pi^2(8k_0+8k_1-1) \nonumber\\ 
& &-4\pi(h_1+6\gamma^2-\gamma)
+4(t_2-u_1-2\gamma h_2-6\gamma^3) \\
\rho_6 &=& (4\pi)^3(s_0 + s_1 + s_2 + s_3)-48\pi^3(k_0+k_1+k_2) \nonumber\\
& & +2\pi^3+\pi^2-4\pi h_2+4(t_3-u_2) \label{zeror}
\end{eqnarray*}

\subsection{Expressions of the coefficients appearing in eq.(\ref{aal2})}

\begin{eqnarray*}
\alpha_1 &=& 
3\bigg[\bigg(\ln\frac{\pi^2}{2}-2\gamma\bigg)^3+(4\pi)^3s_3\bigg]\\
\alpha_2 &=& 
- 3\bigg[6\bigg(\ln\frac{\pi^2}{2}\bigg)^3
+(3\pi-36\gamma+5)\bigg(\ln\frac{\pi^2}{2}\bigg)^2 \\ & &
+ 4\Big(18\gamma^2+2h_1-3\gamma\pi-5\gamma-3\Big)\ln\frac{\pi^2}{2} \\ & &
+4\Big(3\gamma^2\pi-12\gamma^3 + 5\gamma^2 - 4\gamma h_1 
+6\gamma - 16\pi^3 s_2 + 48\pi^3s_3 - t_1\Big)\bigg]\\
\alpha_3 &=& 
2\bigg[18\bigg(\ln\frac{\pi^2}{2}\bigg)^3
+ 6(3\pi-18\gamma+5)\bigg(\ln\frac{\pi^2}{2}\bigg)^2 \\ & & 
+\Big(7\pi^2-9\pi(8\gamma-1)+216\gamma^2-120\gamma+48h_1-66\Big)
\ln\frac{\pi^2}{2} \\ & &
+ 96\pi^3(s_1 - 2s_2 + 3s_3) - 14\gamma\pi^2 
+ 6\pi(12\gamma^2-3\gamma+2h_1-3) \\ & &
-6(16\gamma h_1 + 24\gamma^3 - 20\gamma^2 - 22\gamma 
+ 4t_1 - t_2 +1)\bigg]\\
\alpha_4 &=&
- 4\bigg[6\bigg(\ln\frac{\pi^2}{2}\bigg)^3
+ 3(3\pi-12\gamma+5)\bigg(\ln\frac{\pi^2}{2}\bigg)^2 \\ & &
+\Big(7\pi^2-9\pi(4\gamma-1)+72\gamma^2-60\gamma+24h_1-30\Big)
\ln\frac{\pi^2}{2} \\ & &
+96\pi^3(s_1+s_3)-14\gamma\pi^2 
+ 6\pi(6\gamma^2-3\gamma+2h_1-3) \\ & &
-3(16\gamma h_1 + 16\gamma^3 - 20\gamma^2 - 20\gamma 
+ 4t_1 - 2t_2 + u_2 +1)\bigg]
\end{eqnarray*}

\subsection{Expressions of the coefficients appearing in eq.(\ref{tilb4})}

\begin{eqnarray}
\chi_1 &=& 
\bigg(\ln\frac{\pi^2}{2}-2\gamma\bigg)^3+(4\pi)^3s_3\nonumber\\
\chi_2 &=& 
6\bigg(\ln\frac{\pi^2}{2}\bigg)^3 
+ (3\pi - 36\gamma+1)\bigg(\ln\frac{\pi^2}{2}\bigg)^2 \nonumber\\ & &
+ 4[12k_1\pi^2+ 18\gamma^2 -\gamma(3\pi+1)+h_1]\ln\frac{\pi^2}{2} 
+ (4\pi)^3(s_2 + 3s_3) 
\nonumber\\ & &
- 2[\ln(4\pi)-\gamma] 
\cdot\bigg[3\bigg(\ln\frac{\pi^2}{2}\bigg)^2+ 3(\pi- 4\gamma+1)
\ln\frac{\pi^2}{2} \nonumber\\ & &
+ 12\gamma^2 
-6\gamma(\pi+1) + 24\pi^2k_1 + 6h_1 -1\bigg] + 10\ln(4\pi) 
\nonumber\\ & &
- 48\gamma^3 
+ 4\gamma^2(3\pi+1) 
- 2\gamma(48k_1\pi^2+4h_1+5)
+ 4t_1 \nonumber\\
\chi_3 &=& 
3\bigg(\ln\frac{\pi^2}{2}\bigg)^3
- 6(3\gamma+\pi)\bigg(\ln\frac{\pi^2}{2}\bigg)^2 \nonumber\\ & &
+2[3\pi^2(8k_0+8k_1-1)+\pi(12\gamma-1)+18\gamma^2+2h_2]
\ln\frac{\pi^2}{2} 
\nonumber\\ & &
-[\ln(4\pi)-\gamma]\cdot
\bigg[3\bigg(\ln\frac{\pi^2}{2}-2\gamma\bigg)^2 \nonumber\\ & &
+3\pi^2(16k_0+16k_1-1) 
-6\pi+12h_2-8\bigg] + 4\ln(4\pi)
\nonumber\\ & &
+16\pi^3(4s_1 + 8s_2 + 12s_3-3k_1) 
-12\gamma\pi^2(8k_0+8k_1-1) \nonumber\\ & & 
-4\pi(6\gamma^2-\gamma+h_1)
- 24\gamma^3-4\gamma(2h_2+1)+4t_2 \nonumber\\
\chi_4 &=& 4u_2\nonumber
\end{eqnarray}

\end{appendix}

\end{document}